\documentclass[11pt,a4paper,twoside]{article}
\usepackage{cite}
\usepackage{graphicx}
\usepackage{epsfig}
\usepackage{dcolumn}
\usepackage{bm}

\begin{document}

\def \d {{\rm d}}
\def \t {{\Theta}}
\def \k {{\kappa}}
\def \l {{\lambda}}
\def \s {{\sigma}}
\def \tr {{\tilde\rho}}
\def \tv {{\tilde v}}
\def \tz {{\tilde z}}
\def \e {{\epsilon}}
\def \P {{p_\lambda}}
\def \Q {{p_\nu}}

\newcommand{\be}{\begin{equation}}
\newcommand{\ee}{\end{equation}}

\newcommand{\beqn}{\begin{eqnarray}}
\newcommand{\eeqn}{\end{eqnarray}}
\newcommand{\AdS}{anti--de~Sitter }
\newcommand{\AAdS}{\mbox{(anti--)}de~Sitter }
\newcommand{\AAN}{\mbox{(anti--)}Nariai }
\newcommand{\AS}{Aichelburg-Sexl }
\newcommand{\pa}{\partial}
\newcommand{\pp}{{\it pp\,}-}
\newcommand{\ba}{\begin{array}}
\newcommand{\ea}{\end{array}}

\def \bl {\mbox{\boldmath{$\ell$}}}
\def \bn {\mbox{\boldmath{$n$}}}
\def \bm #1 {\mbox{\boldmath{$m^{(#1)}$}}}
\def \bk {\mbox{\boldmath{$k$}}}

\title{Black rings with a small electric charge: gyromagnetic ratios and algebraic alignment}
\author{Marcello Ortaggio\thanks{ortaggio`AT'science.unitn.it}  
\\ Dipartimento di Fisica, Universit\`a degli Studi di Trento, \\
	and INFN, Gruppo Collegato di Trento, Via Sommarive 14, 38050 Povo (Trento), Italy 
 \\ \\
and \\ \\
 	Vojt\v ech Pravda\thanks{pravda@math.cas.cz} \\ Mathematical Institute, Academy of Sciences, \v{Z}itn\'{a} 25, 115 67 Prague
1, Czech Republic}

\date{\today}
\maketitle

\abstract{We study electromagnetic test fields in the background of vacuum black rings using Killing vectors as vector potentials. We consider both spacetimes with a rotating $S^1$ 
and with a rotating $S^2$ and we demonstrate, in particular, that the gyromagnetic ratio of slightly charged black rings takes the value $g=3$ (this will in fact apply to a wider class of spacetimes). We also observe that a $S^2$-rotating black 
ring immersed in an external ``aligned'' magnetic field completely expels the magnetic flux in the extremal limit. 
Finally, we discuss the mutual alignment of principal null directions of the Maxwell 2-form and of the Weyl tensor, 
and the algebraic type of exact charged black rings. In contrast to spherical black holes, charged rings display 
new distinctive features and provide us with an explicit example of algebraically general (type G) spacetimes in 
higher dimensions. Appendix~\ref{app_S2} contains some global results on black rings with a rotating 2-sphere. Appendix~\ref{app_gyrom} shows that $g=D-2$ in any $D\ge 4$ dimensions for test electromagnetic fields generated by a time translation.} 

\vspace{.2cm}






\section{Introduction}
\label{sec_introduction}

While the study of black holes in higher dimensions has been for a long time motivated by supergravity and 
string theory, recent extra-dimension scenarios have raised further interest in view of possible new observable 
phenomenology (see, e.g., \cite{Horowitz05,Gibbons05,Kanti04} for reviews and references). 
Important early higher dimensional solutions include extensions to any dimension $D>4$ of the Schwarzschild, 
Reissner-Nordstr\"{o}m \cite{Tangherlini63} and Kerr black holes \cite{MyePer86}. However, subsequent investigations have 
shown that even at the classical level gravity in higher dimensions exhibits much richer dynamics than in $D=4$. 

One of the most remarkable features of higher dimensions is the non-uniqueness of the rotating spherical black holes of Myers and Perry \cite{MyePer86}. 
In five dimensions, vacuum black rings have been constructed that have a $S^1\times S^2$ event horizon and that can carry (in a certain parameter range) the same mass and spin as the $S^3$ holes of \cite{MyePer86}. 
There exist by now a number of different black ring solutions that are the subject of ongoing research in  general relativity and string theory (cf.~\cite{EmpRea06} and references therein).  
In particular, rotating black rings with electric charge have been found in the $D=5$ Einstein-Maxwell-Chern-Simons theory,
 e.g. the first supersymmetric ring constructed in \cite{Elvangetal04} and the non-supersymmetric solution of \cite{ElvEmpFig05}. On the other hand, corresponding solutions of the 
standard Einstein-Maxwell equations which carry both electric charge and angular momenta (and are asymptotically flat) 
are not known.\footnote{This in fact applies also to spherical black holes, since as yet there is no Kerr-Newman solution 
in $D>4$. In the case of zero angular momentum, static black rings with electric charge in $D=5$ were constructed 
in \cite{IdaUch03} (but be aware of an incorrect expression for the Maxwell field in eq.~(139) therein, see 
also \cite{KunLuc05,Yazadjiev05}) and necessarily contain a conical singularity.}  

In this context, we study  charged black rings in the limit of a small electric charge. Namely, we analyze physical properties of electromagnetic fields that solve Maxwell's equations in the background of vacuum black rings. In order to do that, we will employ the well known fact  that in a {\em vacuum} spacetime, any Killing vector field $\xi^\mu$ can be taken as a vector potential that automatically satisfies the sourceless Maxwell equations (in the Lorentz gauge) \cite{Papapetrou66,Wald74}. This easily follows from contracting the definition of the Riemann tensor applied to $\xi^\mu$, which yields the identity $\xi^{\mu;\nu}_{\ \ \ \; ;\nu}=-R^\mu_{\ \; \nu}\xi^\nu$. 
The corresponding electromagnetic field is to be understood as a {\em test} field on the given spacetime background, in the sense that its contribution to the energy-momentum tensor of the spacetime (which would appear at second order) is neglected. Therefore, the solution physically makes sense only for sufficiently weak Maxwell fields.\footnote{Indeed, such test fields are equivalent to the linear limit of exact Einstein-Maxwell solutions obtained by applying a Harrison transformation \cite{Stephanibook} along the given Killing vector, at least in $D=4$.} 
This property was already used \cite{Wald74} to study electromagnetic fields on the background of stationary, axisymmetric, asymptotically flat spacetimes, including  in particular the Kerr black hole. Remarkably, it was shown in \cite{Wald74} (and in references therein) that in $D=4$ the thus  generated solutions provide the {\em unique} (stationary, axisymmetric) perturbation which adds an electric charge to a vacuum black hole (or puts the hole into a uniform, aligned magnetic field). Although, to our knowledge, a similar uniqueness has not been proven in $D>4$, the same Killing-technique has been recently employed to analyze test fields in the $D=5$ Myers-Perry spacetime \cite{AliFro04}. In this contribution we will generalize some of the results of  \cite{AliFro04}, such as the value of gyromagnetic ratios, to five-dimensional asymptotically flat spacetimes 
with three commuting Killing vectors, and in particular extend them to the specific case of black rings. In addition, we shall analyze algebraic properties of Maxwell fields in the spacetime of $D=5$ black holes and black rings, both in the limit of weak fields and within exact Einstein-Maxwell solutions (when the latter exist, i.e. in static spacetimes).

In section~\ref{sec_asymptotic} we review asymptotic properties of $D=5$ asymptotically flat vacuum spacetimes with three commuting Killing vectors, in particular the definition of conserved quantities. In section~\ref{sec_test} we study test electromagnetic fields in such backgrounds using Killing vectors to construct the vector potential. We thus clarify the interplay between spacetime geometry and physical properties of the Maxwell field such as electric charge, magnetic field components and gyromagnetic ratios. Using suitable asymptotic coordinates, in section~\ref{sec_ring_S1} we analyze in detail the specific case of the rotating black ring of \cite{EmpRea02prl}. We also briefly comment on ``intrinsic'' magnetic dipoles and on the non-uniqueness of charged solutions (as opposed to the $D=4$ results of \cite{Wald74}). Section~\ref{sec_ring_S2} considers test fields in the background of different black rings, which rotate ``along a 2-sphere'' \cite{MisIgu06,Figueras05}. These solutions admit an extremal limit in which an external magnetic flux is completely expelled from the event horizon. 
In a broader context, in section~\ref{sec_algebraic} we discuss the mutual alignment of principal null directions of the Maxwell 2-form and of the Weyl tensor 
in the case of test fields in several black hole/black ring spacetimes. We also consider the more general case of certain static exact Einstein-Maxwell solutions. We present concluding remarks in section~\ref{sec_concl}. Appendix~\ref{app_S2} clarifies certain global properties of the black rings of \cite{MisIgu06,Figueras05}, whereas appendix~\ref{app_algebraic} presents some technical details which complement section~\ref{sec_algebraic}. In appendix~\ref{app_gyrom} we find the value $g=D-2$ for the gyromagnetic ratio of test electromagnetic fields generated by a time translation in any stationary asymptotically flat spacetime with $D\ge 4$ dimensions.

\section{General asymptotics}
\label{sec_asymptotic}

Let us consider a $D=5$ asymptotically flat vacuum spacetime. In addition we assume that the metric admits three commuting Killing vector fields, namely that it is stationary and invariant under rotations in two different planes. 
There exist suitable asymptotic Cartesian coordinates in which the metric functions behave as \cite{MyePer86} ($r^2=x_ix^i$; $i,j,k=1,\ldots,4$) 
\beqn
 & & g_{tt}=-1+\frac{8}{3\pi}\frac{M}{r^2}+O(r^{-3}) , \nonumber \\
 & & g_{ti}=-\frac{4}{\pi}\frac{x^k}{r^4}J^{ki}+O(r^{-4}) , \label{asym_metric_cartesian} \\  
 & & g_{ij}=\left(1+\frac{4}{3\pi}\frac{M}{r^2}\right)\delta_{ij}+O(r^{-3}) , \nonumber 
\eeqn
where $M$ is the mass of the system, and $J^{ki}$ an antisymmetric $4\times 4$ tensor which represents the angular momentum. By a suitable rotation, we can always set the spacelike Killing vectors in the form $x_1\pa_2-x_2\pa_1$ and $x_3\pa_4-x_4\pa_3$, respectively, so that  $J^{ki}$ takes the standard form with only two independent non-vanishing components, given by the spin parameters $J_\psi\equiv-J^{12}$ and $J_\phi\equiv-J^{34}$. 
 Now, for our purposes it is more convenient to replace the above Cartesian coordinates by asymptotic spherical coordinates adapted to the Killing symmetries (cf.~\cite{CohWal72} in $D=4$),
namely we take
\be
 \qquad x_1=r\sin\theta\cos\psi , \quad x_2=r\sin\theta\sin\psi ,  \qquad  x_3=r\cos\theta\cos\phi , \quad
        x_4=r\cos\theta\sin\phi , 
 \label{spherical}
\ee
with $\theta\in[0,\pi/2]$. The three Killing vectors are thus simply given by
\be
 \xi_{(t)}^\mu=\delta^\mu_t , \qquad  \xi_{(\psi)}^\mu=\delta^\mu_\psi , \qquad \xi_{(\phi)}^\mu=\delta^\mu_\phi ,
 \label{killings}
\ee
and the asymptotic behaviour~(\ref{asym_metric_cartesian}) can be rewritten as
\beqn
 & & g_{tt}=-1+\frac{8}{3\pi}\frac{M}{r^2}+O(r^{-3}) , \nonumber \\
 & & g_{t\psi}=\frac{4}{\pi}\frac{\sin^2\theta}{r^2}J_\psi+O(r^{-3}) , \qquad g_{t\phi}=\frac{4}{\pi}\frac{\cos^2\theta}{r^2}J_\phi+O(r^{-3}) , \nonumber \\ 
 & & g_{rr}=1+\frac{4}{3\pi}\frac{M}{r^2}+O(r^{-3}) , \qquad g_{\theta\theta}=r^2\left(1+\frac{4}{3\pi}\frac{M}{r^2}\right)+O(r^{-1}) , \nonumber \label{asym_metric_polar} \\  
 & & g_{\psi\psi}=r^2\sin^2\theta\left(1+\frac{4}{3\pi}\frac{M}{r^2}\right)+O(r^{-1}) , \\ 
 & & g_{\phi\phi}=r^2\cos^2\theta\left(1+\frac{4}{3\pi}\frac{M}{r^2}\right)+O(r^{-1}) , \nonumber \\
 & & g_{tr}=O(r^{-4}) , \quad g_{t\theta}=O(r^{-3}) , \quad g_{r\theta}=O(r^{-2}) , \nonumber \\ 
 & & g_{r\psi}=O(r^{-2})=g_{r\phi} , \quad g_{\theta\psi}=O(r^{-1})=g_{\theta\phi} , \quad g_{\psi\phi}=O(r^{-1}) . \nonumber
\eeqn
In particular, the flat asymptotic metric takes the form
\be
 \d s^2_0=-\d t^2+\d r^2+r^2(\d\theta^2+\sin^2\theta\d\psi^2+\cos^2\theta\d\phi^2) ,
  \label{flat_spherical}  
\ee
and for the metric determinant one finds $\sqrt{-g}=r^3[\sin\theta\cos\theta+O(r^{-2})]$.

Note that the mass and angular momenta defined by the expansions~(\ref{asym_metric_cartesian}) and  (\ref{asym_metric_polar}) can equivalently be computed as Komar integrals over a 3-sphere at spatial infinity (cf.~\cite{MyePer86,CohWal72}). In fact, one can show that 
\be
 \xi_{(t)}^{t;r}=\frac{8}{3\pi}\frac{M}{r^3}+O(r^{-4}) , \qquad      
  \xi_{(\psi)}^{t;r}=\frac{8}{\pi}\frac{\sin^2\theta}{r^3}J_\psi+O(r^{-4}) , \qquad       
  \xi_{(\phi)}^{t;r}=\frac{8}{\pi}\frac{\cos^2\theta}{r^3}J_\phi+O(r^{-4}) .
\ee
It follows that 
\be
 M=-\frac{3}{32\pi}\int_{S^3_\infty}{^*\d\mbox{\boldmath$\xi$}_{(t)}} , \qquad J_\psi=-\frac{1}{16\pi}\int_{S^3_\infty}{^*\d\mbox{\boldmath$\xi$}_{(\psi)}} , \qquad J_\phi=-\frac{1}{16\pi}\int_{S^3_\infty}{^*\d\mbox{\boldmath$\xi$}_{(\phi)}} ,
 \label{komar}
\ee
where (from now on) $\mbox{\boldmath$\xi$}_{(t)}$, $\mbox{\boldmath$\xi$}_{(\psi)}$ and $\mbox{\boldmath$\xi$}_{(\phi)}$ denote the three 1-forms corresponding to the covariant Killing fields~(\ref{killings}).  

The angular momenta can also be related to the asymptotic behaviour of the twist potential of the time-like Killing vector $\mbox{\boldmath$\xi$}_{(t)}$. Indeed, using the expansions~(\ref{asym_metric_polar}) and the orthonormal cotetrad 
\beqn
 \mbox{\boldmath$\omega$}^{(0)}= & & \left[-\left(-1+\frac{8}{3\pi}\frac{M}{r^2}\right)+O(r^{-3})\right]^{1/2}\Bigg[\d t+\left(\frac{\frac{4}{\pi}\frac{1}{r^2}}{-1+\frac{8}{3\pi}\frac{M}{r^2}}+O(r^{-3})\right) \nonumber  \\
 		& & {}\times\left(J_\psi\sin^2\theta\,\d\psi+J_\phi\cos^2\theta\,\d\phi\right)+O(r^{-4})\d r+O(r^{-3})\d\theta\Bigg] , \nonumber \\
 \mbox{\boldmath$\omega$}^{(1)}= & & \left[1+\frac{4}{3\pi}\frac{M}{r^2}+O(r^{-3})\right]^{1/2}\d r , \label{cotetrad} \nonumber \\
 \mbox{\boldmath$\omega$}^{(2)}= & & r\left[1+\frac{4}{3\pi}\frac{M}{r^2}+O(r^{-3})\right]^{1/2}\left[\d\theta+O(r^{-4})\d r+O(r^{-3})(\d\psi+\d\phi)\right] , \\
 \mbox{\boldmath$\omega$}^{(3)}= & & r\left[1+\frac{4}{3\pi}\frac{M}{r^2}+O(r^{-3})\right]^{1/2}\left[\sin\theta\,\d\psi+O(r^{-4})\d r+O(r^{-3})\d\phi\right] , \nonumber \\
 \mbox{\boldmath$\omega$}^{(4)}= & & r\left[1+\frac{4}{3\pi}\frac{M}{r^2}+O(r^{-3})\right]^{1/2}\left[\cos\theta\,\d\phi+O(r^{-4})\d r+O(r^{-3})\d\psi\right] \nonumber ,
\eeqn
one finds that the twist 2-form 
\be
 \mbox{\boldmath$\omega$}={^*}(\mbox{\boldmath$\xi$}_{(t)}\wedge\d\mbox{\boldmath$\xi$}_{(t)}) ,
 \label{twist}
\ee
satisfies
\beqn
\mbox{\boldmath$\omega$}= & & \frac{8}{\pi}\frac{1}{r^4}\left[\mbox{\boldmath$\omega$}^{(1)}\wedge(J_\phi\sin\theta\,\mbox{\boldmath$\omega$}^{(3)}+J_\psi\cos\theta\,\mbox{\boldmath$\omega$}^{(4)})-
      \mbox{\boldmath$\omega$}^{(2)}\wedge(J_\phi\cos\theta\,\mbox{\boldmath$\omega$}^{(3)}-J_\psi\sin\theta\,\mbox{\boldmath$\omega$}^{(4)})\right] \nonumber \\
      & & {}+O(r^{-5}) . 
\eeqn
Therefore, the twist potential one-form (such that $\mbox{\boldmath$\omega$}=\d\mbox{\boldmath$\chi$}$) behaves as 
\be
 \mbox{\boldmath$\chi$}=-\frac{8}{\pi}\frac{\sin\theta}{2r^3}J_\phi\,\mbox{\boldmath$\omega$}^{(3)}-\frac{8}{\pi}\frac{\cos\theta}{2r^3}J_\psi\,\mbox{\boldmath$\omega$}^{(4)}+O(r^{-4}) .
 \label{twist_pot}      
\ee

\section{Test Maxwell fields}

\label{sec_test}

\subsection{General potential: electric charge and uniform magnetic fields}

In the spirit of \cite{Wald74}, we study here properties of test electromagnetic fields in a generic spacetime that 
satisfies the assumptions of section~\ref{sec_asymptotic} (this includes the case considered in \cite{AliFro04}).
We take a general linear combination of the three Killing vectors~(\ref{killings}) as a test vector potential, namely 
\be
 A^{\mu}=\alpha\delta^\mu_{t}+\beta\delta^\mu_{\psi}+\gamma\delta^\mu_{\phi} ,
 \label{test_A}
\ee
with $\alpha$, $\beta$ and $\gamma$ being arbitrary constants. The covariant field strength $\mbox{\boldmath$F$}=\d\mbox{\boldmath$A$}$ is thus
\be
 \mbox{\boldmath$F$}=\alpha\d\mbox{\boldmath$\xi$}_{(t)}+\beta\d\mbox{\boldmath$\xi$}_{(\psi)}+
    \gamma\d\mbox{\boldmath$\xi$}_{(\phi)} ,
 \label{test_F}    
\ee
in which, of course, $\mbox{\boldmath$\xi$}_{(t)}=g_{tt}\d t+g_{t\psi}\d\psi+g_{t\phi}\d\phi$, etc.. 
In particular, using eqs.~(\ref{asym_metric_polar}) and (\ref{cotetrad}) one finds the leading asymptotic terms
\be
  \mbox{\boldmath$F$}=2\beta\left(\sin\theta\,\mbox{\boldmath$\omega$}^{(1)}+\cos\theta\,\mbox{\boldmath$\omega$}^{(2)}\right)\wedge\mbox{\boldmath$\omega$}^{(3)}+
  	2\gamma\left(\cos\theta\,\mbox{\boldmath$\omega$}^{(1)}-\sin\theta\,\mbox{\boldmath$\omega$}^{(2)}\right)\wedge\mbox{\boldmath$\omega$}^{(4)}+O(r^{-1}) .
 \label{Fasympt}
\ee 
This has the form of a uniform magnetic field $\mbox{\boldmath$F$}=B_\psi\d x_1\wedge\d x_2+B_\phi\d x_3\wedge\d x_4$ (recall eq.~(\ref{spherical}))  with two non-vanishing components 
\be
 B_\psi=2\beta , \qquad B_\phi=2\gamma .
 \label{magnetic}
\ee
The axial Killing vectors thus generate a test Maxwell field which asymptotically approaches a uniform magnetic field, whereas the field generated by the timelike Killing vector vanishes at infinity. 

Also, from the standard definition of the electric charge
\be
 Q=\frac{1}{2\pi^2}\int_{S^3_\infty}{^*}\mbox{\boldmath$F$} 
\ee
with eqs.~(\ref{komar}) one immediately finds the relation
\be
 Q=-\alpha\frac{16}{3\pi}M-\beta\frac{8}{\pi}J_\psi-\gamma\frac{8}{\pi}J_\phi ,
 \label{charge}
\ee
for the charge $Q$ added to the original spacetime by the vector potential~(\ref{test_A}).
This demonstrates that, in the presence of angular momenta, also the spacelike generators contribute to the electric charge. 
Inverting eqs.~(\ref{magnetic}) and (\ref{charge}) one can reexpress the potential~(\ref{test_A}) in terms of physical quantities as
\be
A^{\mu}=-\frac{3\pi}{16}\frac{Q}{M}\,\delta^\mu_{t}+\frac{1}{2}B_\psi\left(\delta^\mu_{\psi}-\frac{3J_\psi}{2M}\,\delta^\mu_{t}\right)
    +\frac{1}{2}B_\phi\left(\delta^\mu_{\phi}-\frac{3J_\phi}{2M}\,\delta^\mu_{t}\right) .
 \label{test_A2}
\ee
Note that the timelike Killing vector enters the above expression even in the case $Q=0$. This can be interpreted as an inductive electric field generated by the rotation of the spacetime, and it is in particular responsible for the effect of charge accretion by a black hole placed in a uniform magnetic field \cite{Wald74,AliFro04}. 

\subsection{Induced magnetic dipoles and gyromagnetic ratios}

Let us now concentrate on the case in which the uniform magnetic field components are vanishing, i.e. $B_\psi=0=B_\phi$, so that the asymptotic term~(\ref{Fasympt}) is zero. Then the vector potential~(\ref{test_A2}) is parallel to the timelike Killing vector $\xi_{(t)}^\mu$. Using the expansions~(\ref{asym_metric_polar}) it is straightforward to determine the various covariant components $A_\mu$ to the leading order 
\be
 A_t\approx \frac{3\pi}{16}\frac{Q}{M}-\frac{Q}{2r^2} , \qquad A_\psi\approx -3\frac{QJ_\psi}{2M}\frac{\sin^2\theta}{2r^2} , \qquad A_\phi\approx -3\frac{QJ_\phi}{2M}\frac{\cos^2\theta}{2r^2} ,
 \label{leading_potential}
\ee
while $A_r$ and $A_\theta$ approach zero faster. The additive constant which appears in $A_t$ is only a removable gauge term, so that the leading term in $A_t$ is the standard (five-dimensional) Coulomb potential. On the other hand, from the leading terms in $A_\psi$ and $A_\phi$ one reads off the two magnetic ``dipole moments'' (cf., e.g., the normalization of \cite{AliFro04,Herdeiro00})
\be
 \mu_\psi=3\frac{QJ_\psi}{2M} , \qquad \mu_\phi=3\frac{QJ_\phi}{2M} . 
 \label{dipoles}
\ee
From these, the gyromagnetic ratios 
\be 
 g_\psi=3=g_\phi  
\ee
readily follow (note that the assumptions we have used so far do not distinguish between the two rotating Killing vectors, which explains why $g_\psi=g_\phi$). 

Equivalently, these can be also derived from geometrical properties of the Maxwell field strength. When $B_\psi=0=B_\psi$, this is simply (see eqs.~(\ref{test_F}) and (\ref{charge}))
\be
 \mbox{\boldmath$F$}=-\frac{3\pi}{16}\frac{Q}{M}\,\d\mbox{\boldmath$\xi$}_{(t)} .
\ee
It is now obvious that the electric part of the 3-form dual to the Maxwell field will be proportional to the twist 2-form~(\ref{twist}), i.e.
\be
 {^*}F_{\mu\nu\rho}\xi^\rho_{(t)}={^*}F_{\mu\nu t}=-\frac{3\pi}{16}\frac{Q}{M}\,\omega_{\mu\nu} .
\ee
This implies a similar proportionality relation between the ``magnetic'' potential $\mbox{\boldmath$\varphi$}$ (defined by ${^*}F_{\mu\nu t}=\varphi_{\nu,\mu}-\varphi_{\mu,\nu}$) and the twist potential 1-form $\mbox{\boldmath$\chi$}$
\be
 \mbox{\boldmath$\varphi$}=-\frac{3\pi}{16}\frac{Q}{M}\,\mbox{\boldmath$\chi$} .  
\ee
Hence from eq.~(\ref{twist_pot}) we have
\be
 \mbox{\boldmath$\varphi$}=3\frac{QJ_\phi}{2M}\frac{\sin\theta}{2r^3}\mbox{\boldmath$\omega$}^{(3)}+3\frac{QJ_\psi}{2M}\frac{\cos\theta}{2r^3}\mbox{\boldmath$\omega$}^{(4)}+O(r^{-4}) .
\ee
Using the definitions of \cite{AliFro04} one again finds eqs.~(\ref{dipoles}).

Note that the above conclusions are rather general, since they apply to any vacuum spacetime satisfying the assumptions of section~\ref{sec_asymptotic}, without involving the knowledge of the full metric.\footnote{In $D=4$, the fact that the Kerr-Newman solution in the full Einstein-Maxwell theory has the value $g=2$ has been known for a long time \cite{Carter68}. A ``universality'' of such gyromagnetic ratio was subsequently found for general stationary, axisymmetric, asymptotically flat spacetimes in the case of test fields derived from the time translation Killing vector \cite{Wald74} (cf.~appendix~\ref{app_gyrom} for the $D$-dimensional case), and for the corresponding Einstein-Maxwell exact solutions generated by a Harrison transformation \cite{ReiTre75} (this relies on the fact that such solutions display the same asymptotics to the relevant order). Examples of solutions with a different $g$ are given, for instance, in \cite{CohTioWal73}. See, e.g., \cite{Herdeiro00} for a few references that discuss the gyromagnetic ratio of objects of Kaluza-Klein and string theory.} In particular, the gyromagnetic ratio we have found agrees with the results of \cite{AliFro04}, obtained in the specific case of the $D=5$ Myers-Perry black hole. It is also consistent with \cite{Aliev06prd}, which considered $D\ge 5$ spherical black holes with an arbitrarily large electric charge but with a small angular momentum (see also \cite{Aliev06}). In the $D=5$ Einstein-Maxwell-Chern-Simons theory there exist supersymmetric rotating charged black holes \cite{Breckenridgeetal97} and black rings \cite{Elvangetal04} that also have $g=3$  \cite{Herdeiro00,Elvangetal05}. On the other hand, recent numerical analysis in odd spacetime dimension $D\ge 5$ \cite{KunNavVie06} indicates that $g$ is a function of the conserved charges in the case of Einstein-Maxwell black holes with {\em generic} values of angular momentum and electric charge (but in the limit of small spin/charge the conclusions of \cite{KunNavVie06} agree with the perturbative values of \cite{AliFro04,Aliev06prd}, and in particular $g\to3$ in $D=5$). 

In the next sections we shall confirm our general results in the special case of rotating black rings, and we shall study more specific features of test Maxwell fields in such backgrounds. 

\section{The $S^1$-black ring}
\label{sec_ring_S1}

The first rotating vacuum black ring was obtained in \cite{EmpRea02prl}. In the coordinates of \cite{Emparan04} (and after trivial rescalings), the metric is
\beqn
 \d s^2= & & -\frac{F(y)}{F(x)}\left(\d t+C(\nu,\lambda)L\frac{1+y}{F(y)}\d\psi\right)^2 \nonumber \label{ring} \\
 & & {}+\frac{L^2}{(x-y)^2}F(x)\left[-\frac{G(y)}{F(y)}\d\psi^2-\frac{\d y^2}{G(y)}+\frac{\d x^2}{G(x)}+\frac{G(x)}{
      F(x)}\d\phi^2\right] ,
\eeqn
where
\be
 F(\zeta)=\frac{1+\lambda\zeta}{1-\lambda} , \qquad G(\zeta)=(1-\zeta^2)\frac{1+\nu\zeta}{1-\nu} , \qquad    
 C(\nu,\lambda)=\sqrt{\frac{\lambda(\lambda-\nu)(1+\lambda)}{(1-\nu)(1-\lambda)^3}} .
 \label{FG}
\ee

The parameters satisfy $0<\nu\le\lambda<1$ and $L>0$. As for the coordinate range, we take $y\in(-\infty,-1]$, $x\in[-1,+1]$. To avoid conical singularities at the axes $x=-1$ and $y=-1$, the angular coordinates must have the standard periodicity 
\be
 \Delta\phi=2\pi=\Delta\psi .
 \label{range}
\ee
The ring is balanced (i.e., conical singularities are absent also at $x=+1$) if
\be
 \lambda=\frac{2\nu}{1+\nu^2} .
 \label{equilibrium}
\ee 
At $y\to-\infty$ the spacetime has a inner spacelike curvature singularity, $y=-1/\nu$ is a horizon and $y=-1/\lambda$ an ergosurface, both with topology $S^1\times S^2$. The algebraic type of the Weyl tensor of the black ring solution is $I_i$ \cite{PraPra05}. The spacetime~(\ref{ring}) is asymptotically flat near spatial infinity $x,y\to -1$.

In order to study asymptotic properties, it is convenient to introduce suitable new asymptotic coordinates $(r,\theta)$ by the substitution (meaningful, say, for $r\gg L$) 
\beqn
 & & x=-1+\frac{2L^2}{r^2}\cos^2\theta+\frac{4L^4}{r^4}\cos^2\theta(k_1\cos^2\theta+k_2) , \nonumber \label{asympt_coords} \\
 & & y=-1-\frac{2L^2}{r^2}\sin^2\theta+\frac{4L^4}{r^4}\sin^2\theta(k_1\sin^2\theta+k_2) , 
\eeqn
with parameters
\be
 k_1=-\frac{-9\nu+5\lambda\nu+3+\lambda}{2(1-\lambda)(1-\nu)} , \qquad  
 k_2=\frac{3\lambda\nu-6\nu+2+\lambda}{2(1-\lambda)(1-\nu)} ,
\ee
and $\theta\in[0,\pi/2]$. Asymptotic spacelike infinity thus corresponds to {$r\to \infty$}, where the metric~(\ref{ring}) approaches the Minkowskian line element~(\ref{flat_spherical}), and the metric coefficients behave as 
\beqn
 & & g_{tt}=-1+\frac{2\lambda}{1-\lambda}\frac{L^2}{r^2}+O(r^{-4}) , \qquad g_{t\psi}= C(\nu,\lambda)\sin^2\theta\frac{2L^3}{r^2}+O(r^{-4}) , \nonumber \\
 & & g_{rr}=1+\frac{\lambda}{1-\lambda}\frac{L^2}{r^2}+O(r^{-4}) , \qquad g_{r\theta}=O(r^{-3}) , \nonumber \label{asym_metric} \\  
 & & g_{\theta\theta}=r^2\left(1+\frac{\lambda}{1-\lambda}\frac{L^2}{r^2}\right)+O(r^{-2}) , \\
 & & g_{\psi\psi}=r^2\sin^2\theta\left(1+\frac{\lambda}{1-\lambda}\frac{L^2}{r^2}\right)+O(r^{-2}) , \nonumber \\ 
 & & g_{\phi\phi}=r^2\cos^2\theta\left(1+\frac{\lambda}{1-\lambda}\frac{L^2}{r^2}\right)+O(r^{-2}) , \nonumber
\eeqn
the remaining ones being identically zero. 
By comparison with eq.~(\ref{asym_metric_polar}), one reads off the mass and angular momentum of the black ring
\be
 M=\frac{3\pi L^2}{4}\frac{\lambda}{1-\lambda} , \qquad J_\psi=\frac{\pi L^3}{2}\,C(\nu,\lambda) . 
 \label{mass}
\ee
Obviously $J_\phi=0$ and the ring rotates only along the $S^1$ parametrized by $\psi$.

In order to construct test fields on the black ring spacetime, we proceed as in section~\ref{sec_test}. Taking a contravariant vector potential of the form~(\ref{test_A}), one now has the only non-vanishing covariant components
\be
 A_t=\alpha g_{tt}+\beta g_{t\psi} , \qquad A_\psi=\alpha g_{t\psi}+\beta g_{\psi\psi} , \qquad A_\phi=\gamma g_{\phi\phi} ,
 \label{potential_S1}
\ee 
with the metric coefficients given in eq.~(\ref{ring}). 

\subsection{Black ring with electric charge}

The $g_{\psi\psi}$ and $g_{\phi\phi}$ components do not vanish near infinity, and give raise to an external magnetic field via eq.~(\ref{potential_S1}) (cf. section~\ref{sec_test}). The latter is absent when $\beta=0=\gamma$, in which case the Maxwell field strength has the only non-zero components
\beqn
 & & F_{xt}=\alpha\frac{\lambda(1+\lambda y)}{(1+\lambda x)^2} , \qquad F_{yt}=-\alpha\frac{\lambda}{1+\lambda x} , \nonumber \label{maxwell} \\
 & & F_{x\psi}=\alpha C(\nu,\lambda)L\lambda(1-\lambda)\frac{1+y}{(1+\lambda x)^2} , \qquad F_{y\psi}=-\alpha C(\nu,\lambda)L\frac{1-\lambda}{1+\lambda x} , 
\eeqn
and it vanishes asymptotically in suitable coordinates, since the expansion~(\ref{asym_metric}) determines the following behavior of the potential for $r\to\infty$
\beqn
 & & A_t=-\alpha+\alpha\frac{2\lambda}{1-\lambda}\frac{L^2}{r^2}+O(r^{-4}) , \nonumber \\
 & & A_\psi=\alpha C(\nu,\lambda)\sin^2\theta\frac{2L^3}{r^2}+O(r^{-4}) , \label{asym_potential} \\
 & & A_\phi=0 \nonumber .
\eeqn
These are a special subcase of eq.~(\ref{leading_potential}) with electric charge and dipole moment given by 
\be
 Q=-4\alpha L^2\frac{\lambda}{1-\lambda} , \qquad \mu_\psi=-4\alpha C(\nu,\lambda)L^3 .
\ee
Recalling also eq.~(\ref{mass}), this explicitly shows that that the gyromagnetic ratio of the rotating black ring of \cite{EmpRea02prl} with a small charge is $g_\psi=3$, as expected from the general conclusions of section~\ref{sec_test}. This coincides with the results of \cite{Elvang03} for a rotating black ring in heterotic supergravity (in the small charge limit and up to a different normalization), and of \cite{Elvangetal05} for supersymmetric black rings. However, it is worth emphasizing that (as opposed to the corresponding situation in $D=4$ \cite{Wald74}) we have not proven any uniqueness result for the above solution~(\ref{maxwell}). In other words, while eq.~(\ref{maxwell}) 
certainly provides a solution (regular in the exterior region and on the horizon, and vanishing at infinity) which adds a charge $Q$ to the black ring~(\ref{ring}), there might exist different solutions which add the same charge (perhaps with a different gyromagnetic ratio).\footnote{Similar comments apply to the charged black holes of \cite{AliFro04}.} Indeed, we shall mention explicit examples in the next subsection. The algebraic structure of the Maxwell field~(\ref{maxwell}) (in the static case) will be discussed in section~\ref{sec_algebraic}. 

\subsection{Black rings with dipole charge and their non-uniqueness} 

The method based on Killing vectors employed above clearly does not lead to a general solution of the Maxwell equations, which would require a study beyond the scope of the present work. We just present here a different solution with interesting physical properties, especially in view of the discussion concluding the preceding subsection. One can easily show that taking $A_t=0=A_\psi$ and
\be
 A_\phi=c_0+c_1(1+x) , 
 \label{dipole}
\ee 
with $c_0$ and $c_1$ arbitrary constants, provides a solution of the Maxwell equations on the background~(\ref{ring}) without any unphysical singularities in the field strength (e.g., at the horizon). This describes the field of an intrinsic magnetic dipole with ``local charge'' \cite{Emparan04}
\be
 q\equiv\frac{1}{4\pi}\int_{S^2}\mbox{\boldmath$F$}=c_1 ,
\ee
where $\mbox{\boldmath$F$}=c_1\d x\wedge\d\phi$ is the Maxwell 2-form (locally) defined by $\mbox{\boldmath$F$}=\d \mbox{\boldmath$A$}$, 
and $S^2$ indicates a 2-sphere parametrized by $(x,\phi)$ at constant $t$, $y$, and $\psi$. In fact, such a test electromagnetic field coincides with the weak-field limit of the solution of the full Einstein-Maxwell equations (a dipole black ring) presented in \cite{Emparan04}. Asymptotically, it behaves as
\be
 A_\phi=c_0+c_1\frac{2L^2}{r^2}\cos^2\theta+O(r^{-4}) , 
 \label{asym_dipole}
\ee
i.e. essentially as the component $A_\psi$ in eq.~(\ref{asym_potential}), which represents a magnetic dipole induced by the rotation of the electric charge.

Thanks to the linearity of Maxwell's equations, one can now arbitrarily superimpose solutions of the form~(\ref{potential_S1}) (e.g., with $\beta=0=\gamma$) and (\ref{dipole}). In particular, since the constant $c_1$ is an arbitrary (``small'') real number, one can describe an infinity of black rings with fixed electric charge (together, of course, with mass and angular momentum) but with an arbitrary value of the dipole  (note, however, that eq.~(\ref{maxwell}), i.e. the solution with to $c_1=0$, is still the unique solution within such a family that adds a charge $Q$ and has zero dipole field in the static limit, so that this non-uniqueness does not affect the value $g_\psi=3$ discussed above). In the special case $Q=0$ (i.e., $\alpha=0$ but $c_1\neq0$), this is a ``linearized'' version of the continuous non-uniqueness observed in \cite{Emparan04} for dipole black rings with zero charge in the full Einstein-Maxwell(-dilaton) theory. 

\subsection{Black ring in an external magnetic field}

We have above analyzed test fields corresponding to a charged (dipole) black ring with a zero background magnetic field. Let us now just mention the complementary situation of a neutral black ring immersed in an external ``uniform'' magnetic field, which arises when $Q=0$ and corresponds to the choice (cf.~eqs.~(\ref{charge}) and (\ref{mass})) 
\be
 \alpha=-\sqrt{\frac{(\lambda-\nu)(1+\lambda)}{\lambda(1-\nu)(1-\lambda)}}\beta L
\ee
in eq.~(\ref{potential_S1}). Note that exact solutions to the full Einstein-Maxwell theory representing (dipole) black rings in an external magnetic (Melvin) field have been obtained in \cite{Ortaggio05} via a suitable Harrison transformation (in the case of spin along $\psi$ and $\mbox{\boldmath$A$}=A_\phi\d\phi$).

\section{The $S^2$-black ring}
\label{sec_ring_S2}

Vacuum black rings with a rotating 2-sphere have been presented in \cite{MisIgu06} and \cite{Figueras05} using different coordinates (cf.~\cite{IguMis06} for the relating coordinate transformation). The metric of \cite{Figueras05} can be written as 
\beqn
 \d s^2 & & = -\frac{H_2(x,y)}{H_1(x,y)}\left(\d t-C(\sigma,\lambda)L\frac{y(1-x^2)}{H_2(x,y)}\d\phi\right)^2+\frac{L^2}{(x-y)^2}H_1(x,y) \nonumber \label{ring_S2} \\
  & & \hspace{-.3cm} {}\times\left[(y^2-1)\frac{F(x)}{H_1(x,y)}\d\psi^2+\frac{1}{F(y)}\frac{\d y^2}{y^2-1}+\frac{1}{F(x)}\frac{\d x^2}{1-x^2}+(1-x^2)\frac{F(y)}{H_2(x,y)}\d\phi^2\right] ,
\eeqn
where
\beqn
 & & H_1(x,y)=\frac{1+\lambda x+(\sigma xy)^2}{1-\lambda+\sigma^2} , \qquad H_2(x,y)=\frac{1+\lambda y+(\sigma xy)^2}{1-\lambda+\sigma^2} , \nonumber \\  
 & & F(\zeta)=\frac{1+\lambda\zeta+(\sigma\zeta)^2}{1-\lambda+\sigma^2} , \qquad 
 C(\sigma,\lambda)=\frac{\sigma\lambda}{(1-\lambda+\sigma^2)^{3/2}} , \label{HF} \\
 & & \mbox{with } \quad 2\sigma<\lambda<1+\sigma^2, \qquad \sigma\in(0,1) \nonumber .
\eeqn
The coordinate $(x,y)$ range as $x\in[-1,+1]$ $y\in(-\infty,-1]\cup[1,\infty)$ (see appendix~\ref{app_S2}). 
The above metric is regular at the axes $x=-1$ and $y=-1$ provided eq.~(\ref{range}) holds. With this choice, however, there is a conical singularity at $x=+1$ representing a disk-shaped membrane inside the ring. There exist an outer and an inner horizon at $y_{\pm}=(-\lambda\pm\sqrt{\lambda^2-4\sigma^2})/(2\sigma^2)$, which in the extremal limit $\lambda=2\sigma$ coincide at $y_h=-1/\sigma$. The spacetime has a curvature singularity at $x=0$ with $y=\pm\infty$.
We refer to appendix~\ref{app_S2} for further details, including the localization of ergosurfaces. In the exterior region, the spacetime~(\ref{ring_S2}) is asymptotically flat at $x,y\to -1$. Using again the asymptotic coordinates~(\ref{asympt_coords}), but now with the parameters
\be
 k_1=\frac{5\lambda-3+\sigma^2}{2(1-\lambda+\sigma^2)} , \qquad  
 k_2=\frac{2-3\lambda-2\sigma^2}{2(1-\lambda+\sigma^2)} ,
\ee
one finds the expansion for {$r\to\infty$} of the only non vanishing metric components
\beqn
 & & g_{tt}=-1+\frac{2\lambda}{1-\lambda+\sigma^2}\frac{L^2}{r^2}+O(r^{-4}) , \qquad g_{t\phi}= -2C(\sigma,\lambda)\cos^2\theta\frac{2L^3}{r^2}+O(r^{-4}) , \nonumber \\
 & & g_{rr}=1+\frac{\lambda}{1-\lambda+\sigma^2}\frac{L^2}{r^2}+O(r^{-4}) , \qquad g_{r\theta}=O(r^{-3}) , \nonumber \label{asym_metric_S2} \\  
 & & g_{\theta\theta}=r^2\left(1+\frac{\lambda}{1-\lambda+\sigma^2}\frac{L^2}{r^2}\right)+O(r^{-2}) , \\
 & & g_{\psi\psi}=r^2\sin^2\theta\left(1+\frac{\lambda}{1-\lambda+\sigma^2}\frac{L^2}{r^2}\right)+O(r^{-2}) , \nonumber \\ 
 & & g_{\phi\phi}=r^2\cos^2\theta\left(1+\frac{\lambda}{1-\lambda+\sigma^2}\frac{L^2}{r^2}\right)+O(r^{-2}) . \nonumber
\eeqn
Recalling eq.~(\ref{asym_metric_polar}), the mass and angular momentum of the black ring follow 
\be
 M=\frac{3\pi L^2}{4}\frac{\lambda}{1-\lambda+\sigma^2} , \qquad J_\phi=-\pi L^3C(\sigma,\lambda) . 
 \label{mass_S2}
\ee

Again, test fields on the above spacetime can be constructed as in sections~\ref{sec_test} and \ref{sec_ring_S1}. Starting from eq.~(\ref{test_A}), one obviously obtains 
\be
 A_t=\alpha g_{tt}+\gamma g_{t\phi} , \qquad A_\psi=\beta g_{\psi\psi} , \qquad A_\phi=\alpha g_{t\phi}+\gamma g_{\phi\phi} , 
 \label{potential_S2}
\ee 
with the metric coefficients as in eq.~(\ref{ring_S2}). 

\subsection{Black ring with electric charge}

Adding an electric charge with no external magnetic field amounts to having $\beta=0=\gamma$, in which case the Maxwell field is explicitly given by
\beqn
 & & F_{xt}=\alpha\lambda\frac{1+\lambda y+\sigma^2xy^2(2y-x)}{[(1+\lambda x+(\sigma xy)^2]^2} , \qquad F_{yt}=-\alpha\lambda\frac{1+\lambda x+\sigma^2x^2y(2x-y)}{[(1+\lambda x+(\sigma xy)^2]^2} , \nonumber \label{maxwell_S2} \\
 & & F_{x\phi}=-\alpha\frac{\sigma\lambda L}{\sqrt{1-\lambda+\sigma^2}}\frac{y(\lambda+2x+\lambda x^2+2\sigma^2xy^2)}{[(1+\lambda x+(\sigma xy)^2]^2} , \\  & & F_{y\phi}=\alpha\frac{\sigma\lambda L}{\sqrt{1-\lambda+\sigma^2}}\frac{(1-x^2)[1+\lambda x-(\sigma xy)^2]}{[(1+\lambda x+(\sigma xy)^2]^2} . \nonumber 
\eeqn
The $r\to\infty$ fall-off is determined by eq.~(\ref{asym_metric_S2}) as 
\beqn
 & & A_t=-\alpha+\alpha\frac{2\lambda}{1-\lambda+\sigma^2}\frac{L^2}{r^2}+O(r^{-4}) , \nonumber \\
 & & A_\psi=0 , \label{asym_potential_S2} \\
 & & A_\phi=-2\alpha C(\sigma,\lambda)\cos^2\theta\frac{2L^3}{r^2}+O(r^{-4}) \nonumber ,
\eeqn
corresponding to the following values of the electric charge and dipole moment 
\be
 Q=-4\alpha L^2\frac{\lambda}{1-\lambda+\sigma^2} , \qquad \mu_\phi=8\alpha C(\sigma,\lambda)L^3 .
\ee
Similarly as in the previous section, this demonstrates that the gyromagnetic ratio is ${g_\phi=3}$ for a slightly charged black ring with a rotating 2-sphere. Note also that the asymptotic form of $A_\phi$ is identical to that of a dipole field, cf.~eq.~(\ref{asym_dipole}). In the present case, however, there is no associated local magnetic charge (essentially because $A_\phi$ is regular over each $S^2$). 

\subsection{Extremal black ring in a magnetic field: flux expulsion}

The complementary situation of a black ring with $Q=0$ under the influence of an external magnetic field occurs when one takes
\be
 \alpha=\frac{2\sigma L}{\sqrt{1-\lambda+\sigma^2}}\gamma 
\ee
in the components~(\ref{potential_S2}). In this case, it is interesting to write down the explicit form of the potential, i.e.
\beqn
 A_t= & & -\gamma\frac{\sigma L}{\sqrt{1-\lambda+\sigma^2}}\frac{2+\lambda y+x^2y(\lambda+2\sigma^2y)}{1+\lambda x+(\sigma xy)^2} , \nonumber \\
 A_\psi= & & \beta\frac{L^2}{1-\lambda+\sigma^2}\frac{y^2-1}{(x-y)^2}[1+\lambda x+(\sigma x)^2] , \\
 A_\phi=& & \gamma\frac{\sigma^2\lambda L^2}{1-\lambda+\sigma^2}\frac{y(1-x^2)}{1+\lambda y+(\sigma xy)^2}
 							\frac{2+\lambda y+x^2y(\lambda+2\sigma^2y)}{1+\lambda x+(\sigma xy)^2} \\ \nonumber
 				& & {}+\gamma\frac{L^2}{1-\lambda+\sigma^2}\frac{1-x^2}{(x-y)^2}\frac{1+\lambda x+(\sigma xy)^2}{1+\lambda y+(\sigma xy)^2}
 									(1+\lambda y+\sigma^2y^2) .
\eeqn  
From now on, we shall be concentrating on {\em extremal} black rings, defined by $\lambda=2\sigma$ (an upper index ``ex'' will denote quantities evaluated at extremality). From the above equations, one can easily verify that in this case the $A^{\mbox{\tiny ex}}_t$ and $A^{\mbox{\tiny ex}}_\phi$ components are proportional to $(1+\sigma y)$. This seems to indicate that they vanish at the degenerate horizon $y=y_h=-1/\sigma$. However, the coordinates presently in use become singular precisely there. One therefore needs new coordinates that extend through $y=-1/\sigma$, for instance the set $(v,x',y',\psi',\chi)$ defined by (cf. also, e.g., \cite{Elvangetal04})
\beqn
 & & \d t=\d v+\left[\frac{A_1\sigma}{1+\sigma y}+\frac{A_2\sigma^2}{(1+\sigma y)^2}\right]\d y , \nonumber \\
 & & \d\phi=\d\chi+\left[\frac{B_1\sigma}{1+\sigma y}+\frac{B_2\sigma^2}{(1+\sigma y)^2}\right]\d y , \\
 & & x=x' , \qquad y=y' , \qquad \psi=\psi' , \nonumber  
\eeqn
with the constant parameters
\beqn
 & & A_1=\frac{2\sigma L}{1-\sigma}(B_1+\sigma B_2) , \qquad A_2=\frac{2\sigma L}{1-\sigma}B_2 , \nonumber \\
 & & B_1=\frac{\sigma^2}{(1+\sigma)\sqrt{1-\sigma^2}} , \qquad B_2=\frac{\sqrt{1-\sigma^2}}{\sigma(1+\sigma)} .
\eeqn
In the new coordinate system, the vector potential takes the form $A^{\mbox{\tiny ex}}=A^{\mbox{\tiny ex}}_v\d v+A^{\mbox{\tiny ex}}_{y'}\d y'+A^{\mbox{\tiny ex}}_{\psi'}\d\psi'+A^{\mbox{\tiny ex}}_\chi\d\chi$, with
\beqn
 A^{\mbox{\tiny ex}}_v & = & A^{\mbox{\tiny ex}}_t=-\gamma\frac{2\sigma L}{1-\sigma}\frac{1+\sigma x^2y}{1+2\sigma x+(\sigma xy)^2}(1+\sigma y) , \nonumber \\
 A^{\mbox{\tiny ex}}_{\psi'} & = & A^{\mbox{\tiny ex}}_\psi=\beta\frac{L^2}{(1-\sigma)^2}\frac{y^2-1}{(x-y)^2}(1+\sigma x)^2 , \nonumber \\
 A^{\mbox{\tiny ex}}_{y'} & = & A^{\mbox{\tiny ex}}_t\left[\frac{A_1\sigma}{1+\sigma y}+
 					\frac{A_2\sigma^2}{(1+\sigma y)^2}\right]+A^{\mbox{\tiny ex}}_\phi\left[\frac{B_1\sigma}{1+\sigma y}+\frac{B_2\sigma^2}{(1+\sigma y)^2}\right] \nonumber \\
 					& = & \left[-4\sigma^2\frac{(1+\sigma x^2y)^2}{1+2\sigma x+(\sigma xy)^2}+\frac{1-x^2}{(x-y)^2}[1+2\sigma x+(\sigma xy)^2]\right] \label{Aex} \\ 
 					& & {}\times\gamma\frac{\sigma L^2}{(1-\sigma)^2}\frac{\sigma B_2+B_1(1+\sigma y)}{1+2\sigma y+(\sigma xy)^2}-\gamma\frac{4\sigma^4L^2B_2}{(1-\sigma)^2}\frac{1+\sigma x^2y}{1+2\sigma x+(\sigma xy)^2} , \nonumber  \\
 A^{\mbox{\tiny ex}}_\chi & = & A^{\mbox{\tiny ex}}_\phi=\gamma\frac{4\sigma^3L^2}{(1-\sigma)^2}\frac{y(1-x^2)}{1+2\sigma y+(\sigma xy)^2}
 							\frac{1+\sigma x^2y}{1+2\sigma x+(\sigma xy)^2}(1+\sigma y) \nonumber \\ 
 				& & \hspace{1cm} {}+\gamma\frac{L^2}{(1-\sigma)^2}\frac{1-x^2}{(x-y)^2}\frac{1+2\sigma x+(\sigma xy)^2}{1+2\sigma y+(\sigma xy)^2}
 									(1+\sigma y)^2 . \nonumber
\eeqn

The amount of magnetic flux $\Phi$ across a portion of the horizon ($y'=y=-1/\sigma$) bounded by a closed curve $\Gamma$ is given by the line integral $\Phi=\oint_\Gamma A$. It is thus obvious from eq.~(\ref{Aex}) that the only contribution to $\Phi$ will come from $A^{\mbox{\tiny ex}}_{\psi'}$. In particular, for $\beta=0$ (i.e., $B_\psi=0$) the flux is completely expelled from the horizon of extremal black rings. This is analogous to the flux expulsion  (``Meissner effect'') from extremal Kerr black holes in $D=4$, observed for aligned uniform magnetic fields (through the ``upper half'' of the horizon) in \cite{KinLasKun75} and generalized to any stationary axisymmetric magnetic field (across any part of the horizon) in \cite{BicJan85} (non-aligned fields are not expelled \cite{BicJan85}). See \cite{BicDvo80} for the discussion of flux expulsion from extremal {R}eissner-{N}ordstr{\"{o}}m black holes in $D=4$, \cite{AliFro04} for the case of $D=5$ extremal Myers-Perry black holes, and  \cite{ChaEmpGib98} for similar effects in Kaluza-Klein and string theories.

\section{Algebraic properties of test Maxwell fields}

\label{sec_algebraic}

In this section we analyze algebraic properties of a test field $\mbox{\boldmath$F$}=\d \mbox{\boldmath$A$}$ 
in the case of
a potential
\be
  A^\mu=\alpha\delta^\mu_t 
\ee 
parallel to the timelike Killing vector $\partial_t$ of certain static or stationary black hole spacetimes. 
Study of null eigenvectors $\bk$ (satisfying $F^\mu_{\ \, \nu}k^\nu=\lambda_{(k)} k^\mu$) of the test field \mbox{\boldmath$F$} generated by Killing vectors 
provides a link between symmetries and algebraic structure
of the spacetime. As a remarkable example of this interplay, it was shown in \cite{Mars:CQG1999} that in four 
dimensions the Kerr metric is the only  asymptotically flat vacuum spacetime with a Killing vector 
$\mbox{\boldmath $\xi$}$  approaching a time translation at infinity,
for which
the null eigenvectors of the 2-form $\mbox{\boldmath$F$}=\d \mbox{\boldmath $\xi$}$   coincide with the principal null directions
 of the Weyl tensor.
In fact,  since the full solution of the Einstein-Maxwell system representing a charged rotating black hole is known 
(the Kerr-Newman solution) it can be shown that also the null eigenvectors of the exact field strength tensor 
\mbox{\boldmath $\tilde F$} coincide  with the principal null directions of the Weyl tensor.  
(Note that for the Kerr-Newman metric the vector $\bk$ in the Kerr-Schild form, which is necessarily a principal null direction, does not depend on charge.)
 
Here we want to analyze possible alignment of  principal null directions of a vacuum spacetime with  
eigenvectors of a test field \mbox{\boldmath$F$} 
 in the case of five dimensional Schwarzschild-Tangherlini and Myers-Perry black holes
 and of static black rings. We shall comment also on  alignment of the {\em exact} 
field strength tensor \mbox{\boldmath $\tilde F$} for the corresponding solutions of the 
full Einstein-Maxwell equations, when these are known.\footnote{Note that, as a consequence of Einstein's equations with $4\pi T_{\mu\nu}=\tilde F^\rho_{\ \, \mu}\tilde F_{\rho\nu}-\frac{1}{4}\tilde F^{\rho\sigma}\tilde F_{\rho\sigma}g_{\mu\nu}$, eigenvectors of \mbox{\boldmath $\tilde F$} are also eigenvectors of the Ricci tensor.}

In order to do that, in following calculations we will use the frame
$$
\bm{0} 
= \bn, \qquad \bm{1} 
=\bl, \qquad \bm{i} 
 , \qquad\qquad  i= 2 \dots 4,
$$
with two null vectors   $\bn,\ \bl$  
\be
\ell^\mu  \ell_\mu = n^\mu  n_\mu  = 0, \qquad   \ell^\mu  n_\mu  = 1, \qquad\qquad   \mu = 0  \dots 4,
\label{framecond1}
\ee
 and three spacelike vectors $\bm{i} $ 
\be
 m^{(i)\mu }m^{(j)}_\mu =\delta_{ij}, \qquad  m^{(i)\mu }\ell_\mu =0=m^{(i)\mu }n_\mu ,  \qquad\qquad  i,j = 2 \dots 4. 
\label{framecond2}
\ee
Then, we will employ the recently developed generalization of the Petrov classification of the Weyl tensor in higher dimensions \cite{Coleyetal04,Milsonetal05}. For the Maxwell field, one can similarly use its decomposition into different boost order components \cite{Milson:2004wr} (see also \cite{Hall} for detailed information in the four-dimensional case), i.e. $F_{\mu\nu}=2F_{0i}n_{[\mu} m^{(i)}_{\nu]}+2F_{01}n_{[\mu}\ell_{\nu]}+F_{ij}m^{(i)}_{[\mu}m^{(j)}_{\nu]}+2F_{1i}\ell_{[\mu} m^{(i)}_{\nu]}$, in order to define aligned null directions. Let us note that in all five-dimensional cases studied below \mbox{\boldmath$F$} and \mbox{\boldmath $\tilde F$} possess two null eigendirections and are thus of the algebraically special type $(1,1$) \cite{Milson:2004wr} with all their non-vanishing components having boost order zero (that is, $F_{0i}=0=F_{1i}$). In this sense one could also regard \mbox{\boldmath$F$} and \mbox{\boldmath $\tilde F$} as being of type D.

\subsection{Schwarzschild-Tangherlini}
Let us analyze alignment of \mbox{\boldmath$F$} and \mbox{\boldmath $\tilde F$} with the Weyl tensor in the
case of five-dimensional static vacuum and charged black holes \cite{Tangherlini63}. 

\subsubsection{The vacuum case (test field)}
The five-dimensional spherically symmetric vacuum
Schwarzschild-Tangherlini black hole in hyperspherical coordinates has the form
\be
{\rm d}s^2 = -f^2 {\rm d} t^2 + f^{-2} {\rm d} r^2  + r^2 {\rm d} \Omega_3^2,
\label{ST-metric}
\ee
where
\be
f^2=1-\frac{\mu}{r^2},
\ee
and ${\rm d} \Omega_3^2$ is the line element of the unit 3-sphere. 

The  null eigenvectors and corresponding eigenvalues of the mixed test field strength $F^\mu_{\ \, \nu}$ are 
\beqn
 \bl & = & {\partial_t}  - f^2 {\partial_r}: \quad \lambda_{(\ell)}=\frac{2 \alpha \mu}{r^3} , \label{ST-l} \\
 \bn & = & {\partial_t} + f^2  {\partial_r}: \quad \lambda_{(n)}=-\frac{2 \alpha \mu}{r^3} .
\label{ST-n}
\eeqn
In the 3-space of spacelike eigenvectors corresponding to the eigenvalue 0, we  choose the following orthonormal basis
\be
\bm{2} =  \frac{1}{r} {\partial_\theta} , \ \ \ 
\bm{3} = \frac{1}{r \sin \theta} {\partial_\phi}, \ \ \
\bm{4} = \frac{1}{r \cos \theta} {\partial_\psi} ,
\label{ST-m}
\ee
so that $\bl,\ \bn, \bm{2} ,\ \bm{3} $ and $\bm{4} $ satisfy eqs.~(\ref{framecond1}) and (\ref{framecond2}).

In this frame all components of the Weyl tensor with boost orders 2, 1, -1 and -2 (see \cite{Coleyetal04}) vanish. 
Consequently $\bl$ and $\bn$ are also principal null directions (or Weyl aligned null directions - WANDs) of the (type D \cite{Coleyetal04,ColPel06}) 
Weyl tensor and \mbox{\boldmath$F$} is thus completely aligned with the 
Weyl tensor.

\subsubsection{The charged case (full Einstein-Maxwell)}
A charged static five-dimensional black hole is described by the metric~(\ref{ST-metric}) with
\be
f^2=1-\frac{\mu}{r^2} + \frac{e^2}{r^4} ,
\ee
and by the vector potential
\be
\mbox{{\boldmath $\tilde A$}}= -\sqrt{\frac{3}{4}} \frac{e}{r^2} {\rm d} t.
\ee
The null eigenvectors of the field strength tensor $\mbox{\boldmath$\tilde F$}=\d \mbox{\boldmath$\tilde A$}$ are again given by eqs.~(\ref{ST-l}) and (\ref{ST-n}) and are aligned with the type D Weyl tensor.

\subsection{Myers-Perry}
The metric of a five-dimensional vacuum black hole with two rotational parameters $a$, $b$ can be written in the form  \cite{MyePer86} 
\beqn
 {\rm d}s^2=&&-{\rm d}t^2 + \frac{\rho^2 r^2}{ \Delta} {\rm d}r^2 + \rho^2 {\rm d} \theta^2  + (r^2+a^2) \sin^2 \theta
{\rm d} \phi^2 + (r^2+b^2) \cos^2 \theta {\rm d} \psi^2  \\
&& + \frac{\mu}{\rho^2} ({\rm d}t + a \sin^2 \theta {\rm d} \phi
+ b \cos^2 \theta {\rm d} \psi)^2 ,
\eeqn
where
$$
\rho^2=r^2+a^2 \cos^2 \theta + b^2 \sin^2 \theta , \qquad
\Delta=(r^2+a^2)(r^2+b^2)-\mu r^2 .
$$
The two null eigenvectors of the test field \mbox{\boldmath$F$}  
\beqn
L_{\pm} = \frac{(r^2+a^2)(r^2+b^2)}{\Delta} \left( \partial_t -\frac{a}{r^2+a^2} \partial_{\phi} - \frac{b}{r^2+b^2} \partial_{\psi}
  \right) \pm \partial_r ,
\eeqn
corresponding to the eigenvalues
\be
\mp \frac{  2 \alpha\mu r}{\left(r^2+a^2\cos^2 \theta + b^2 \sin^2 \theta \right)^2} ,
\ee
coincide with the two WANDs (of this type D spacetime) identified in \cite{FroSto03,Pravdaetal04}.

An exact Einstein-Maxwell solution corresponding to the charged version of the Myers-Perry metric is unknown. Nevertheless, in view of the above results it seems plausible to expect that for such a solution \mbox{\boldmath $\tilde F$} will be again aligned with the Weyl tensor and that the spacetime will be again of type D. Note, however, that the supersymmetric supergravity black holes of \cite{Breckenridgeetal97} are of type I$_i$ \cite{ColPel06}.

\subsection{Static black rings}

\subsubsection{The vacuum case (test field)}

The metric of a static vacuum black ring is given by eq.~(\ref{ring}) with $\nu=\lambda$.
The Maxwell field $\mbox{\boldmath$F$}$, see eq.~(\ref{maxwell}), admits two null eigenvectors $\bl\equiv L_+$, $\bn\equiv L_-$
\be
L_{\pm} = \pm \frac {\sqrt {- \left( 1+\lambda\,x \right)  \left( x-y \right) 
 \left( x+x\lambda\,y+\lambda+y \right) }L}{ 
 \left( y^2-1 \right)  \left( 1+\lambda\,y \right)  \left( x-y \right) }
\partial_t  
+ {\frac {x^2-1}{y^2-1}}
\partial_x
+ \partial_y ,
\ee
which correspond to the eigenvalues 
\be
 \mp {\frac {\sqrt {- \left( 1+\lambda\,x \right)  \left( x-y \right) 
 \left( x+x\lambda\,y+\lambda+y \right) } \left( x-y \right) \alpha\,
\lambda}{ \left( 1+\lambda\,x \right) ^{2}L}}.
\ee
We complete the frame by
\be
\bm{2} = \frac{1+\lambda x}{1+\lambda y} \partial_x + \partial_y,  \ \ \ 
\bm{3} = \partial_{\phi},  \ \ \ \bm{4} = \partial_{\psi}. 
\ee
Note that this  frame is not normalized but this does not affect further results.

In this frame 
\be
C_{\mu\nu\rho\sigma} {m^{(2)\mu} } \ell^\nu m^{(2)\rho} \ell^\sigma = C_{\mu\nu\rho\sigma} m^{(2)\mu} n^\nu m^{(2)\rho} n^\sigma =
\frac{\lambda L^2}{2}
{\frac { \left( x+x\lambda\,y+\lambda+y \right) }
{ \left( {y}^{2}-1 \right)^3   \left( 1+\lambda\,y \right) ^{2}}}.
\ee
These quantities do not vanish and thus
$\bl$ and $\bn$ are not aligned with the Weyl tensor (which is of type I$_i$ \cite{PraPra05}).
Since even the test field  $\mbox{\boldmath$F$}$ does not preserve principal null directions of the spacetime,
one would {\em a fortiori} expect an analogous situation in the case of an exact Einstein-Maxwell charged static black ring, 
and thus one would also intuitively expect this spacetime to be of algebraically general type G.\footnote{A similar situation arises in $D=4$ when one puts a  Schwarzschild black hole (of type D) into a magnetic Melvin universe (also of type D but with ``different'' principal null directions), which results in the Schwarzschild-Melvin spacetime (of algebraically general type I \cite{BosEst81}).} This turns out to be true (see the next subsection). 

\subsubsection{The charged case  (full Einstein-Maxwell)}
\label{SCBR}

The metric of a static charged black ring is \cite{IdaUch03,KunLuc05,Yazadjiev05}
\be
 \d s^2=-\Lambda^{-2}\frac{F(y)}{F(x)}\d t^2
 +\frac{\Lambda L^2}{(x-y)^2}F(x)\bigg[\frac{1}{F(y)}\frac{\d y^2}{y^2-1}+\frac{1}{F(x)}\frac{\d x^2}{1-x^2}+(1-x^2)\d\phi^2 + (y^2-1)\d\psi^2\bigg] ,
 \label{ring_el}
\ee
with $F(\xi)$ as in eq.~(\ref{FG}) and $\Lambda$ and the electromagnetic potential given by (we take $e^2<1$)
\be
 \Lambda=\frac{1-e^2\frac{F(y)}{F(x)}}{1-e^2} , \qquad \mbox{\boldmath $\tilde A$}=-\frac{\sqrt{3}\,e}{2\Lambda}\frac{F(y)}{F(x)}\,\d t .
\ee

We want to prove that the metric (\ref{ring_el}) is indeed of type G, i.e. that it does not possess WANDs.
Consequently, we have to show that the necessary and sufficient conditions for 
type I (see \cite{Milsonetal05,PraPra05})
\be
I_{\tau\mu\sigma\iota}=\ell^\nu \ell^\rho \ell_{[\tau}C_{\mu]\nu\rho[\sigma}\ell_{\iota]}=0 \label{PNDeq}
\ee
do not admit any real null solution $\bl$.

Let us denote the covariant components of $\bl$ as
\be
\ell_\mu  = (\alpha, \beta, \gamma, \delta, \epsilon).
\ee
The condition that $\bl$ is null leads to a second order polynomial equation in $\alpha, \dots, \epsilon$.
Components of the Weyl tensor for the metric~(\ref{ring_el}) are very complicated and thus eqs.~(\ref{PNDeq}) in general lead to fourth order polynomial equations with  thousands of terms.
Fortunately, by choosing the right sequence of equations, one can  show that eqs.~(\ref{PNDeq}) do not 
admit a real null solution. Let us now outline this procedure. 

Two of the equations (\ref{PNDeq}) are particularly simple -
$ I_{tx\phi \psi} $ implies 
\be
\lambda (e^2-1) \alpha \beta \delta \epsilon = 0 \label{PNDeq1}
\ee
and $ I_{tx\phi \psi} $ leads to
\be
\lambda  e^2 (e^2-1) \alpha \gamma \delta  \epsilon = 0. \label{PNDeq2}   
\ee

First note that for a non-zero $\bl$, the component $\ell_t=\alpha$ must be non-zero  since $\bl$ is null and $t$ is the timelike coordinate.
As a consequence of the previous two equations either $\beta = \gamma=0$ or $\delta \epsilon=0$. In appendix~\ref{app_algebraic}
we study in detail all possibilities compatible with (\ref{PNDeq1}) and (\ref{PNDeq2}) and show that they do not
admit a real null solution. These possibilities are 1) $\beta = \gamma = 0$;  2) $\epsilon = \beta = 0$; 3) $\epsilon = \gamma = 0$;
4) $\epsilon = \delta = 0$; 5) $\delta = \beta = 0$; 6) $\delta = \gamma = 0$; 7) $\delta = 0$, the rest is non-zero; 
8) $\epsilon = 0$, the rest is non-zero.

Consequently, charged static black rings do not admit WANDs  and, 
to our knowledge, 
they provide the first explicitly identified known solutions of type G.
\footnote{In \cite{ColPel06} it was pointed out that the supersymmetric black ring of \cite{Elvangetal04} is not 
more special than type I$_i$, and that it is possibly of type G.}

\section{Conclusions}

\label{sec_concl}

We have studied various features of five-dimensional stationary and (bi-)axisymmetric asymptotically flat spacetimes. 
First, we have discussed properties of test Maxwell fields obtained from Killing generators in vacuum backgrounds. 
Some of these, such as gyromagnetic ratios, depend only on the asymptotics and not on the details of the spacetime. 
In particular, we have  recovered the $g=3$ result of \cite{AliFro04} for Myers-Perry black holes and extended it to more general 
solutions, including both families of vacuum black rings. Then we have elucidated facts that are related to the specific spacetimes 
one considers, for example flux expulsion from extremal black rings. Our results are natural extensions of previous investigations 
in $D=4$, but we have also pointed out important differences. These are mainly related on the one hand to the non-uniqueness of
 perturbative charged solutions in $D=5$, and
on the other hand to the mutual alignment properties of principal null directions of the Maxwell strength and of the Weyl tensor. 
While spherical $D=5$ black holes behave, in this sense, like the $D=4$ Kerr metric (which is in fact uniquely characterized 
by its alignment properties \cite{Mars:CQG1999}), black rings present a more complex structure. 
As an interesting consequence, we have pointed out that exact Einstein-Maxwell solutions describing (static) charged black 
rings have a Weyl tensor of the most general algebraic type (type G). This seems to be the first explicit example of such a 
spacetime. As a side result, in appendix~\ref{app_S2} we have analyzed the global structure of the $S^2$-black ring. 

Although our contribution has been mainly focused on test electromagnetic fields, some of the conclusions extend to exact solutions of theories describing gravity coupled to antisymmetric forms, and may thus stimulate further studies. For example, one could naturally wonder whether exact Einstein-Maxwell black rings with arbitrary electric and dipole charge exist that generalize the solutions of \cite{Emparan04} and that reduce to the test fields of section~\ref{sec_ring_S1} for small $Q$ and $q$. Apart from supersymmetric black rings \cite{Elvangetal04,BenWar04,Elvangetal05,GauGut05b} (in which mass and charge(s) saturate a BPS bound and are therefore related), black ring solutions of the Einstein-Maxwell-Chern-Simons theory exist which contain both electric and dipole charge \cite{ElvEmpFig05}, but these are not independent parameters. Nevertheless, some of our test field results can be recovered in the weak field limit of such solutions (since the Chern-Simons ``correction'' affects Maxwell's equations only at second order). Moreover, the study of alignment of the test field with the Weyl tensor for the Myers-Perry metric in five dimensions seems to indicate that an exact Einstein-Maxwell charged black hole (if found) could
 be again of the type D, with the only non-vanishing components of the field strength tensor and Ricci tensor being of boost order zero. These investigations are left for future work.

\section*{Acknowledgments}

Part of the work of M.O. was carried out at the Institute of Theoretical Physics of the Charles University in Prague,
 with a post-doctoral fellowship from Istituto Nazionale di Fisica Nucleare (bando n.10068/03). V.P. acknowledges support from research plan No AV0Z10190503 and research grant KJB1019403.

\appendix

\section{On the global structure of the $S^2$-black ring}
\renewcommand{\theequation}{A\arabic{equation}}
\setcounter{equation}{0}

\label{app_S2}

In this appendix we discuss some properties of the spacetime~(\ref{ring_S2}), in order to complement (and in part recover) the analysis of \cite{Figueras05} and \cite{MisIgu06,IguMis06} (the latter being in different coordinates). 

\subsection{Metric signature}

Let us first determine regions where the metric~(\ref{ring_S2}) admits Lorentzian signature.

The metric admits five real eigenvalues $\lambda_1 \dots \lambda_5$, where  $\lambda_1=g_{xx}$, $\lambda_2=g_{yy}$,
$\lambda_3=g_{\psi \psi}$.  Expressions for $\lambda_4$ and $\lambda_5$ are quite complicated, but their product
$\lambda_4 \lambda_5=g_{tt} g_{\phi \phi} - g_{t \phi}^2$ is much simpler. One can then use the signs of $(\lambda_4 \lambda_5)$  and $g_{tt}$ to determine the signature of the 2-metric over the subspace $(t,\phi)$, since $\d s^2_{(t,\phi)}=g_{tt}\d t^2+2g_{t\phi}\d t \d\phi+g_{\phi\phi}\d\phi^2=g_{tt}(\d t+g_{t\phi}g_{tt}^{-1}\d\phi)^2+(\lambda_4 \lambda_5)g_{tt}^{-1}\d\phi^2$. 
Now, in order to study the behaviour of the eigenvalues $\lambda_1 \dots \lambda_5$ explicitly we analyze the functions $F(y)$, $H_2(x,y)$, $F(x)$ and $H_1(x,y)$ entering the metric~(\ref{ring_S2}).

The quadratic function $F(y)$ has the minimum for $y=-\lambda / 2 \sigma^2 < -1 $. The value
of $F(y)$ at this point is $(1-\lambda^2 / 4 \sigma^2)/(1-\lambda+\sigma^2)<0.$ On the other hand, $F(-1)=1>0$. Consequently
$F(y)$ has two roots 
\be
 y_{\pm}=\frac{-\lambda\pm\sqrt{\lambda^2-4\sigma^2}}{2\sigma^2} , \qquad y_-< y_+< -1 , 
 \label{horizons}
\ee 
and it is thus strictly positive everywhere except for $y \in [y_-,y_+]$. As an immediate consequence, in the set $x\in(-\infty,-1]\cup[1,\infty)$ with $y\in(-\infty,y_-)\cup(y_+,\infty)$ one finds that necessarily $H_2(x,y)>0$, and in the set $x\in[-1,1]$ with $y\in(y_-,y_+)$ necessarily $H_2(x,y)<0$. 
{ However, the full region of the plane $(x,y)$ with $H_2(x,y)>0$ is given by the semiplane $y>-1/\lambda$ and by the subset of the semiplane $y<-1/\lambda$ that satisfies $x^2>-(1+\lambda y)/(\sigma y)^2$.} 

Of course $F(x)$ and $H_1(x,y)$ present the same behaviour as $F(y)$ and $H_2(x,y)$ up to interchanging $x\leftrightarrow y$. Namely, $F(x)>0\Leftrightarrow x\in(-\infty,y_-)\cup(y_+,\infty)$ (note that, in particular, $F(x)>0$ for $x\in[-1,1]$), and  $H_1(x,y)>0$ in the semiplane $x>-1/\lambda$ and in the region $x<-1/\lambda$ with $y^2>-(1+\lambda x)/(\sigma x)^2$ (consequently, $H_1(x,y)>0$ for $x\in[-1,1]$ with $y\in(-\infty,-1]\cup[1,\infty)$).
 
\subsubsection{Asymptotically flat Lorentzian region}

Let us for the moment concentrate on the special case $x\in(-1,1)$ (we shall return to more general results soon).

From the above observations, one finds that $\lambda_1 = {-L^2 H_1(x,y)}/[{(x-y)^2  (x^2-1) F(x)}]$ is obviously { always}
 positive for 
 $y \in (-\infty,-1) \cup (1,\infty)$, while $\lambda_2= L^2 H_1(x,y) /[(x-y)^2 (y^2-1) F(y)]$ is positive 
for $y \in (-\infty,y_-) \cup (y_+,-1) \cup (1,\infty)$ and negative for $y \in (y_-,y_+)$.
 
Then, $\lambda_3=L^2 (y^2-1) F(x) /(x-y)^2$ is { always} positive for $y \in (-\infty,-1) \cup (1,\infty)$.

The product $\lambda_4 \lambda_5=L^2 (x^2-1) F(y) /(x-y)^2$ is negative for $y \in (-\infty,y_-) \cup (y_+,\infty)$
 and positive for $y \in (y_-,y_+)$. In the latter case, it follows that both $\lambda_4>0$ and $\lambda_5>0$ since $g_{tt}>0$ there, so that the signature of $\d s^2_{(t,\phi)}$ is $(+,+)$ and the overall five-dimensional signature in this region is Lorentzian.

Note finally that for $y \in (-1,1)$ one has both $\lambda_1 \lambda_2<0$ and $\lambda_4 \lambda_5<0$. Consequently,
the signature cannot be Lorentzian (nor Euclidean) there.  

In summary, we have shown { explicitly} that for $x \in (-1,1)$ the signature is Lorentzian for all values of the 
parameters $\sigma, \lambda$ (within the range given in eq.~(\ref{HF})) for $y \in (-\infty,y_-) \cup(y_-,y_+)\cup(y_+,-1)\cup(1,\infty)$, and {\em not}
 Lorentzian for $y \in (-1,1)$. It will become clear in subsection~\ref{asymptS2} that the
 Lorentzian region represents a spacetime that is asymptotically flat at $x,y\to\pm1$. 
Recall however the presence of a conical singularity at $x=+1$ (cf.~section~\ref{sec_ring_S2}).

\subsubsection{Other Lorentzian regions and Euclidean sectors}

\begin{figure}
\begin{center}
\includegraphics*[height=5cm]{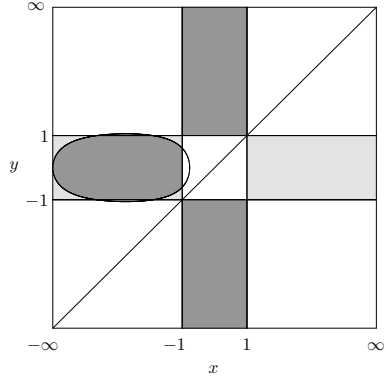}
\end{center}
\caption{Schematic plot of regions of the $(x,y)$ plane where the metric~(\ref{ring_S2}) has Lorentzian (dark grey) and 
Euclidean (light grey) signatures. Note that an appropriate compactification using the tangent function of $y$ (vertical axis) and $x$ (horizontal axis) is used, which enables us to plot their full range. Namely the left[right] edge represents $x=\mp\infty$ and the lower[upper] edge $y=\mp\infty$.
The ``ellipse-like'' curve (defined by $H_1(x,y)=0$) represents a curvature singularity. For definiteness, the picture is drawn with a parameter choice $\lambda>1$.}
 \label{fig_euclid}
\end{figure}

A general study (i.e., dropping the restriction $x\in(-1,1)$) of the signature of the line element~(\ref{ring_S2}) is straightforward but lengthy. We sketch here basic ideas (to be combined with the discussion of the sign of the structural functions presented above) and the main results. It is useful to preliminarily consider the sign of the products $(\lambda_1\lambda_2)$, $(\lambda_4\lambda_5)$ together with the sign of $\lambda_3$ in order to rule out a number of combinations, since these three signs must contain just one minus for the signature being Lorentzian (this is not sufficient, though). Then, more detailed analysis shows that in the case $\lambda_1\lambda_2<0$ one has two Lorentzian regions, i.e. $A: x\in(-1,1)$, $y\in(y_-,y_+)$ and $B: x\in(y_-,y_+)$, $y\in(-1,1)$. When $\lambda_4\lambda_5<0$ one finds another region $C: x\in(-1,1)$, $y\in(-\infty,y_-)\cup(y_+,-1)\cup(1,\infty)$. Finally, in the case $\lambda_3<0$ there is a Lorentzian sector $D$ in the subset of the region $x\in(-\infty,y_-)\cup(y_+,-1)$, $y\in(-1,1)$ defined by $y^2<-(1+\lambda x)/(\sigma x)^2$. Lorentzian signature is not possible elsewhere. To summarize, $A\cup C$ corresponds to the Lorentzian region $x\in(-1,1)$, $y \in (-\infty,y_-) \cup(y_-,y_+)\cup(y_+,-1)\cup(1,\infty)$ discussed previously. This turns out to be disjoint from the third set $B\cup D$, which is non-asymptotically flat since it does not allow $x,y\to-1$ simultaneously (see subsection~\ref{asymptS2}) and is free from conical singularities (with the periodicities of section~\ref{sec_ring_S2}). Lorentzian regions are depicted in Fig.~\ref{fig_euclid}.

To conclude, we just mention that there exist also {\em Euclidean regions} with signature $(-----)$, given by $x\in(1,\infty)$, $y\in(-1,1)$ and by  the subset of $x\in(-\infty,y_-)\cup(y_+,-1)$, $y\in(-1,1)$ defined by $y^2>-(1+\lambda x)/(\sigma x)^2$. Some of these regions are asymptotically flat, see again Fig.~\ref{fig_euclid}.

\subsection{Curvature singularity, asymptotic flatness and horizons}

\label{asymptS2}

In the rest of this appendix we shall focus only on the asymptotically flat Lorentzian region  $A\cup C$ (shaded in Fig.~\ref{fig_ergo}), therefore restricting to the coordinate range
\be
 x\in(-1,1), \quad y\in(-\infty,-1)\cup(1,\infty) .
 \label{domain}
\ee

The Kretschmann scalar $K=R_{\mu\nu\rho\sigma} R^{\mu\nu\rho\sigma}$ has the form
\be
 K=\frac{(x-y)^4 P(x,y)}{{H_1^6(x,y)}},
\ee
where $P(x,y)$ is a polynomial of degree 8 with respect to $y$ and contains the parameters $L$, $\lambda$ and $\sigma$.
Since $H_1(x,y)>0$ in the Lorentzian domain~(\ref{domain}), the denominator of $K$ is nonzero there (note, however, that it can vanish in other regions, cf.~Fig.~\ref{fig_euclid}).   
The limit of $K$ for $y \rightarrow \pm \infty$ is 
\be
\lim_{y \rightarrow  \pm \infty} K= \frac {6 \lambda^2 (\sigma^2 - 2 \lambda x - 2) }{\sigma^6 x^6}.
\ee
 This implies that there exist curvature singularities located at $(x,y)=(0,\pm \infty)$. 

A necessary condition for the spacelike and null infinity for an asymptotically 
flat spacetime is $K=0$ and these can thus be located  at $y=x$, { that is at $x,y\to\pm 1$ in the coordinate range considered here}
 { (this can be seen also from the limit of the metric~(\ref{ring_S2}) \cite{Figueras05})}.

Note that we can identify $y=-\infty$ with $y=+\infty$ and regard the ``upper'' region $y>1$ as analytical continuation of the region
$y<-1$. This  can be seen by performing the coordinate transformation $Y=1/y$ which brings the two regions together. The resulting
metric is Lorentzian for $x \in (-1,1)$, $Y \in (-1,1)$. The curvature singularity is located at $x=0$, $Y=0$ and the rest
of the $Y=0$ line in the $x\in(-1,1)$ interval is regular.  (Analogously we can also join two Euclidean regions by identifying $x=-\infty$ with $x=\infty$, cf.~Fig.~\ref{fig_euclid}).  

Horizons are located at zeros of $F(y)$, i.e. $y=y_\pm<-1$ given by eq.~(\ref{horizons}), and they have topology $S^1\times S^2$ \cite{Figueras05}.

\subsection{Ergosurfaces}

Note that $g_{tt}\to-1$ for both $x,y\to-1$ and $x,y\to+1$. Therefore, in both asymptotic regions $t$ corresponds to a 
timelike coordinate  of an observer at infinity. 
On the other hand, $g_{tt}>0$ in the region $y_-<y<y_+$ between the inner and outer horizon and also in ergoregions outside the outer horizon and inside the inner horizon. These are bounded by ergosurfaces at $g_{tt}=0$ (i.e., $H_2(x,y)=0$).
The solution of this equation is
\be
 x^2=x^2_{\mbox{\tiny e}}(y)=-\frac{1+\lambda y}{\sigma^2y^2} ,
 \label{ergo}
\ee
which defines a real ergosurface provided $0<x^2_{\mbox{\tiny e}}<1$. 
For next purposes it is convenient to reexpress the latter in terms of the coordinate $y$, which defines two disjoint branches, i.e. an {\em outer} ($y^{\mbox{\tiny e}}_+$) and an {\em inner} ($y^{\mbox{\tiny e}}_-$) ergosurface
\be
 y^{\mbox{\tiny e}}_{\pm}(x)=\frac{-\lambda\pm\sqrt{\lambda^2-4\sigma^2 x^2}}{2\sigma^2x^2} .
\ee 
Note that $y^{\mbox{\tiny e}}_+\ge y_+$ and $y^{\mbox{\tiny e}}_-\le y_-$, with equalities holding at $x=\pm1$, so that 
the outer[inner] ergosurface coincides with the outer[inner] with horizon at the poles of $S^2$ \cite{Figueras05}. The metric induced on spatial sections of the ergosurfaces now reads
\beqn
 \d s_{\mbox{\tiny e}}^2= & & \frac{L^2H_1(x,y_{\mbox{\tiny e}})}{(x-y_{\mbox{\tiny e}})^2}\left[(y_{\mbox{\tiny e}}^2-1)\frac{F(x)}{H_1(x,y_{\mbox{\tiny e}})}\d\psi^2+(1-x^2)\left(1+\frac{2(1-x^2)\sigma^2y_{\mbox{\tiny e}}^2}{(1-\lambda+\sigma^2)H_1(x,y_{\mbox{\tiny e}})}\right)\d\phi^2\right. \nonumber \label{ergo_metr} \\
  & &  \left.{}+\left(\frac{4\sigma^2x^2y_{\mbox{\tiny e}}^2}{y_{\mbox{\tiny e}}^2-1}\frac{1-\lambda+\sigma^2}{\lambda^2-4\sigma^2x^2}+\frac{1}{F(x)}\right)\frac{\d x^2}{1-x^2}\right] ,
\eeqn
where $y_{\mbox{\tiny e}}$ generically stands for $y^{\mbox{\tiny e}}_\pm(x)$ according to the branch one considers. Here we have used the simplifying identities $1+\lambda y_{\mbox{\tiny e}}=-(\sigma x y_{\mbox{\tiny e}})^2$ and $F(y_{\mbox{\tiny e}})=(1-x^2)\sigma^2 y_{\mbox{\tiny e}}^2/(1-\lambda+\sigma^2)$.

\begin{figure}
\begin{center}
\includegraphics*[height=6cm]{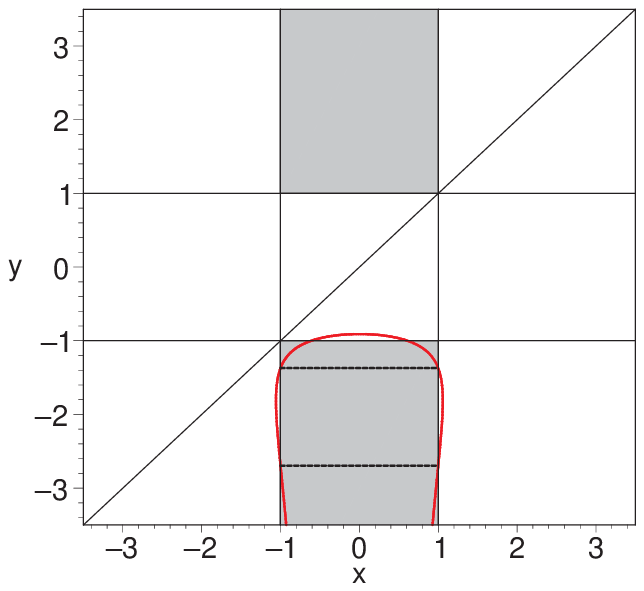}
\hspace{7mm}
\includegraphics*[height=6cm]{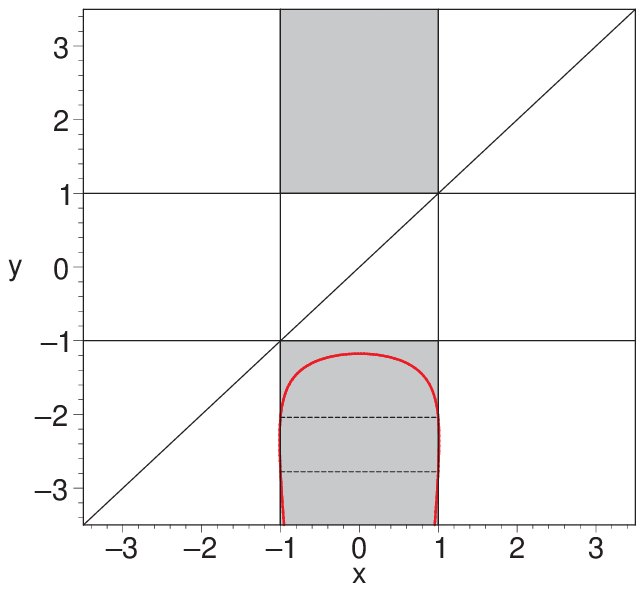}
\end{center}
\caption{Schematic plot of the $S^2$-black ring. Regions of interest with Lorentzian signature are shaded. The line $y=x$,
where the Kretschmann scalar vanishes, is indicated. Horizons
are denoted by dashed lines and ergosurfaces by continuous curves. Fig.~\ref{fig_ergo}.a (left) represents the case with $\lambda>1$ with the ergosurface
``intersecting'' the line $y=-1$. Fig.~\ref{fig_ergo}.b (right) corresponds to the case with $\lambda<1$. See text for details.}
 \label{fig_ergo}
\end{figure}

For further analysis one has  to consider two different possibilities (see Fig.~\ref{fig_ergo}) according 
to the range of the parameters $\lambda$ and $\sigma$, which in general are taken 
to obey $2\sigma<\lambda<1+\sigma^2$ and $\sigma\in(0,1)$. 

For $\lambda>1$ (which is always true if $\sigma\in(\frac{1}{2},1)$), eq.~(\ref{ergo}) always satisfies 
the constraints $0<x^2_{\mbox{\tiny e}}<1$ in regions with $y<y_-$ or $y>y_+$, so that the outer ergosurface extends to any value $y_+\le y^{\mbox{\tiny e}}_+\le -1$ outside the outer horizon, and the inner ergosurface to any $-\infty<y^{\mbox{\tiny e}}_-\le y_-$ inside the inner horizon (cf.~Fig.~~\ref{fig_ergo}.a). 
Interestingly, this implies that the outer ergosurface crosses the axis $y=-1$ at $x=\pm x_0$, with $x_0\equiv\sqrt{\lambda-1}/\sigma$. 
Therefore, in addition to the asymptotically flat region, there is another region with $g_{tt}<0$ ``in the proximity'' 
of the point $x=1$, $y=-1$ and still outside the outer horizon. Thus observers coming from infinity 
have to cross the ergosurface, go through the ergoregion and cross again the ergosurface 
if they want to reach this second region with $g_{tt}<0$ (with no need to cross the horizon).
It also follows that this outer ergosurface consists of two disconnected components described by $x\in(-1,-x_0)$ and $x\in(x_0,1)$, respectively. By inspecting the metric~(\ref{ergo_metr}) it is easy to see that in the coordinate ranges considered here these two components have spherical topology  $S^3$, since the orbits of $\pa_\phi$ close only either at $x=-1$ {\em or} at $x=+1$ on each component, and the orbits of $\pa_\psi$ close at $x=-x_0$ {\em or} at $x=+x_0$ (notice that $(y^{\mbox{\tiny e}}_+)^2-1\sim x^2-x_0^2$ when $x\to\pm x_0$). 
On the other hand, the inner ergosurface inside the inner horizon extends to the curvature singularity at $x=0,$ $y\to-\infty$, and we will not discuss it any further here. 

For $\lambda<1$ (which is possible only if $\sigma\in(0,\frac{1}{2})$) the outer ergosurface is localized in $y_+\le y^{\mbox{\tiny e}}_+\le-1/\lambda$ so that it does not intersect the axis $y=-1$ (cf.~Fig.~~\ref{fig_ergo}.b). It is thus a connected surface of topology $S^1\times S^2$. Notice that this range of parameters includes, in particular, the limit of small rotation $\sigma\to 0$.  

Both the cases considered above admit an extremal limit $\lambda=2\sigma$ (but for $\lambda>1$ this is possible only when $\sigma\in(\frac{1}{2},1)$), in which the two horizons coincide and intersect both the inner and the outer ergosurfaces at $x=\pm1$. 

So far, our discussion has been confined to the physically more interesting case $y<-1$. 
 The asymptotically flat extended region $y>1$ is analogous to the $r<0$ region of the Kerr solution and similarly it has negative mass and contains closed timelike curves (since $g_{\phi\phi}<0$ at $x=0$ for $y\to+\infty$). It does not contain any horizons or ergosurfaces.

\section{Charged black ring is of type G}
\renewcommand{\theequation}{B\arabic{equation}}
\setcounter{equation}{0}

\label{app_algebraic}

Here we study all possibilities compatible with eqs.~(\ref{PNDeq1}) and (\ref{PNDeq2}) as given
in  subsection~\ref{SCBR}.

1) $\beta = \gamma = 0$: We solve $\ell_a \ell^a = 0$ for $\alpha^2$. Then $I_{txy\phi}$ leads to
\be
 \delta \left( \delta^2 \left(y^2-1 \right) + \epsilon^2 \left(1-x^2 \right)   \right) = 0.
\ee 
Since in our region of interest $y^2-1 > 0$ and $1-x^2>0$, the only non-trivial solution is $\delta=0$.
In that case, however, $I_{tyy\psi}$ implies $\epsilon=0$. Thus case 1) does not lead to a non-trivial solution.

2) $\epsilon = \beta = 0$: We solve $\ell_a \ell^a = 0$ for $\alpha^2$. Then $I_{t\phi xy}$ leads to $\delta \gamma=0$.
The case $\gamma=0$ is a subcase of 1). For the case $\delta=0$, $I_{txyx}$ gives $\gamma=0$. 
Thus case 2) does not lead to a non-trivial solution.

3) $\epsilon = \gamma = 0$: We solve $\ell_a \ell^a = 0$ for $\alpha^2$. Then $I_{t\phi xy}$ leads to $\beta \delta =0$.
The case $\beta=0$ is a subcase of 1). For the case $\delta=0$, $I_{tyxy}$ implies $\beta=0$.
Thus case 3) does not lead to a non-trivial solution. 

4) $\epsilon = \delta = 0$: We solve $\ell_a \ell^a = 0$ for $\alpha^2$. Then $I_{t\phi xy}$ leads to $\beta \delta =0$.
Now  $I_{\phi yx \phi}$ and  $I_{\psi y \psi x}$ are two quadratic equations in $\beta$ and  $\gamma$ of the form
\beqn
f_1 \beta^2 + f_2 \beta \gamma + f_3 \gamma^2 &=& 0,\\
f_4 \beta^2 + f_5 \beta \gamma + f_6 \gamma^2 &=& 0,
\eeqn
where $f_i$ are  specific  functions (usually very complicated polynomials or rational functions of the variables $x$ and $y$ 
and parameters $e$, $L$, and $\lambda$ arising in the calculations). In our outline
we do not give the explicit form of these functions but they may be obtained using Maple.
An appropriate linear combination of these equations leads to
\be
\beta (\tilde f_1 \beta + \tilde f_2 \gamma) = 0.
\ee
Now either $\beta=0$ or $\beta=-\gamma \tilde f_2/\tilde f_1 $. 

For the case $\beta=0$, $I_{\phi yx \phi}$ implies $\gamma=0$. For the case $\beta=-\gamma \tilde f_2/\tilde f_1 $,
 $I_{txxy}$ also implies $\gamma=0$. Thus case 4) does not lead to a non-trivial solution. 

5) $\delta = \beta = 0$: We solve $\ell_a \ell^a = 0$ for $\alpha^2$. Then $I_{ytyx}$ implies $\epsilon \gamma =0$.
The case $\epsilon=0$ is a subcase of 4). The case $\gamma=0$ is a subcase of 1). Thus case 5) does not lead to a non-trivial solution. 

6) $\delta = \gamma = 0$: We solve $\ell_a \ell^a = 0$ for $\alpha^2$. Then $I_{txxy}$ implies $\beta \epsilon =0$.
$\beta=0$ is subcase of 1), $\epsilon =0$ is subcase of 3).
Thus case 5) does not lead to a non-trivial solution. 

7) $\delta = 0$, the rest is non-zero: We solve $\ell_a \ell^a = 0$ for $\alpha^2$. Then
 $I_{\psi \phi \psi \phi}$ has the form
\be
f_7 \beta^2 + f_8 \beta \gamma + f_9 \gamma^2 + f_{10} \epsilon^2= 0.
\ee
We solve the above equation for $\epsilon^2$.
Then $I_{yxt\psi}$ and  $I_{yt \psi y}$ lead to 
\beqn
f_{11} \beta^2 + f_{12} \beta \gamma + f_{13} \gamma^2 &=& 0,\\
f_{14} \beta^2 + f_{15} \beta \gamma + f_{16} \gamma^2 &=& 0.
\eeqn
An appropriate linear combination of these two equations is
\be
\gamma (f_{17} \beta + f_{18} \gamma)=0.
\ee
This equation can be solved for $\beta$ but then $I_{tyt\psi}$ implies $\gamma=0$ which belongs to the case 5).
Thus case 7) does not lead to a non-trivial solution. 

8) $\epsilon = 0$, the rest is non-zero: We solve $\ell_a \ell^a = 0$ for $\alpha^2$. Then
 $I_{t x x \phi}$ has the form
\be
f_{19} \beta^2 + f_{20} \beta \gamma + f_{21} \gamma^2 + f_{22} \delta^2 = 0.
\ee
We solve this equation for $\delta^2$ and then $I_{\phi y t y}$ and $I_{\phi y x t}$ reduce to
\beqn
f_{23} \beta^2 + f_{24} \beta \gamma + f_{25} \gamma^2 &=& 0,\\
f_{26} \beta^2 + f_{27} \beta \gamma + f_{28} \gamma^2 &=& 0,
\eeqn
with an appropriate linear combination giving $\gamma=0$. Consequently case 8) does not lead to a non-trivial solution. 

\section{Gyromagnetic ratios in arbitrary dimensions}
\renewcommand{\theequation}{C\arabic{equation}}
\setcounter{equation}{0}

\label{app_gyrom}

The results of section~\ref{sec_test} concerning the value of the gyromagnetic ratio $g$ can be straightforwardly extended to any $D\ge 4$. In this case, an asymptotically flat spacetime is characterized by \cite{MyePer86} 
\beqn
 & & g_{tt}=-1+\frac{16\pi}{(D-2)\Omega_{D-2}}\frac{M}{r^{D-3}}+O(r^{2-D}) , \nonumber \\
 & & g_{ti}=-\frac{8\pi}{\Omega_{D-2}}\frac{x^k}{r^{D-1}}J^{ki}+O(r^{1-D}) , \label{asym_metric_D} \\  
 & & g_{ij}=\left(1+\frac{16\pi}{(D-2)(D-3)\Omega_{D-2}}\frac{M}{r^{D-3}}\right)\delta_{ij}+O(r^{2-D}) , \nonumber 
\eeqn
(which generalizes eq.~(\ref{asym_metric_cartesian})), where $\Omega_{D-2}$ is the area of a unit $(D-2)$-sphere. We now assume that the spacetime is stationary and the corresponding Killing vector field approaches a time translation at infinity, i.e. $\xi_{(t)}^\mu=\delta^\mu_t$. This enables us to construct a suitable test vector potential simply as $A^{\mu}=\alpha\delta^\mu_{t}$, so that $\mbox{\boldmath$A$}=\alpha(g_{tt}\d t+g_{ti}\d x^i)$. A general asymptotically flat spacetime contains $\lfloor (D-1)/2\rfloor$ independent angular momenta in orthogonal planes \cite{MyePer86}. Let $(x^1,x^2)$ be one of these planes. Then we can introduce there asymptotic polar coordinates $(\xi,\psi)$ defined by $x^1=\xi\cos\psi$, $x^2=\xi\sin\psi$ and compute the component of the vector potential along the generator of asymptotic rotations (in the $(x^1,x^2)$ plane) $\pa_\psi=x^1\pa_2-x^2\pa_1$, that is $A_\psi=\alpha\pa_t\cdot\pa_\psi=\alpha(x^1g_{t2}-x^2g_{t1})$. Using the expansion~(\ref{asym_metric_D}) together with adapted coordinates such that $J^{1k}=0$ for $k\neq 2$ and $J^{2k}=0$ for $k\neq 1$ (and $J_\psi\equiv-J^{12}$) one immediately finds near infinity
\be
 A_\psi=\alpha\frac{8\pi J_\psi}{\Omega_{D-2}}\frac{\xi^2}{r^{D-1}} .
 \label{A_D}
\ee

Now, the Komar integral of the 1-form $\mbox{\boldmath$\xi$}_{(t)}$ associated to the Killing vector $\xi_{(t)}^\mu$ gives the mass of the spacetime \cite{MyePer86}
\be
 M=-\frac{1}{16\pi}\frac{D-2}{D-3}\int_{S^{D-2}_\infty}{^*\d\mbox{\boldmath$\xi$}_{(t)}} .
 \label{komar_D}
\ee
Similarly, the electric charge is defined by \cite{MyePer86} 
\be
 Q=\frac{1}{\Omega_{D-2}}\int_{S^{D-2}_\infty}{^*}\mbox{\boldmath$F$}=\frac{\alpha}{\Omega_{D-2}}\int_{S^{D-2}_\infty}{^*\d\mbox{\boldmath$\xi$}_{(t)}} , \label{charge_D} 
\ee
where we have used $\mbox{\boldmath$F$}=\d\mbox{\boldmath$A$}$. From eqs.~(\ref{komar_D}) and (\ref{charge_D}) one finds $-16\pi\alpha M(D-3)=Q\Omega_{D-2}(D-2)$, so that eq.~(\ref{A_D}) can be reexpressed only in terms of physical quantities as
\be
 A_\psi=-(D-2)\frac{QJ_\psi}{2M}\frac{\xi^2}{(D-3)r^{D-1}} .
\ee
With, e.g., the defition of \cite{KunNavVie06} (up to a different normalization of the electric charge) one gets the gyromagnetic ratio
\be
 g_\psi=D-2 .
\ee
This conclusion is in agreement with the value of $g$ found in \cite{Aliev06prd} for slowly rotating Myers-Perry black holes, and with the numerical results of \cite{KunNavVie06} for black holes in odd dimensions $D\ge 5$ if the limit of small charge is taken.



\begin{thebibliography}{10}

\bibitem{Horowitz05}
G.~T. Horowitz, {\it Higher dimensional generalizations of the {K}err black
  hole},  \href{http://xxx.lanl.gov/abs/gr-qc/0507080}{{\tt gr-qc/0507080}}.

\bibitem{Gibbons05}
G.~W. Gibbons, ``Black holes in higher dimensions.'' Talk at the Isaac Newton
  Institute Conference ``Global General Relativity'', Cambridge, August 26,
  2005. Available at {\tt
  http://www.newton.cam.ac.uk/webseminars/pg+ws/2005/gmr/gmrw02/0826/gibbons/}.

\bibitem{Kanti04}
P.~Kanti, {\it Black holes in theories with large extra dimensions: A review},
  {\em Int. J. Mod. Phys. {\rm A}} {\bf 19} (2004) 4899--4951.

\bibitem{Tangherlini63}
F.~R. Tangherlini, {\it Schwarzschild field in $n$ dimensions and the
  dimensionality of space problem},  {\em Nuovo Cimento} {\bf 27} (1963)
  636--651.

\bibitem{MyePer86}
R.~C. Myers and M.~J. Perry, {\it Black holes in higher dimensional
  space-times},  {\em Ann. Phys. (N.Y.)} {\bf 172} (1986) 304--347.

\bibitem{EmpRea06}
R.~Emparan and H.~S. Reall, {\it Black rings}, 
  {\em Class. Quantum Grav.} {\bf 23} (2006) R169--R197.

\bibitem{Elvangetal04}
H.~Elvang, R.~Emparan, D.~Mateos, and H.~S. Reall, {\it A supersymmetric black
  ring},  {\em Phys. Rev. Lett.} {\bf 93} (2004) 211302.

\bibitem{ElvEmpFig05}
H.~Elvang, R.~Emparan, and P.~Figueras, {\it Non-supersymmetric black rings as
  thermally excited supertubes},  {\em JHEP} {\bf 02} (2005) 031.

\bibitem{IdaUch03}
D.~Ida and Y.~Uchida, {\it Stationary {E}instein-{M}axwell fields in arbitrary
  dimensions},  {\em Phys. Rev. {\rm D}} {\bf 68} (2003) 104014.

\bibitem{KunLuc05}
H.~K. Kunduri and J.~Lucietti, {\it Electrically charged dilatonic black
  rings},  {\em Phys. Lett. {\rm B}} {\bf 609} (2005) 143--149.

\bibitem{Yazadjiev05}
S.~S. Yazadjiev, {\it Asymptotically and non-asymptotically flat static black
  rings in charged dilaton gravity},
  \href{http://xxx.lanl.gov/abs/hep-th/0507097}{{\tt hep-th/0507097}}.

\bibitem{Papapetrou66}
A.~Papapetrou, {\it Champs gravitationnels stationnaires \`a sym\'etrie
  axiale},  {\em Ann. Inst. H. Poincar{\'e} {\em A}} {\bf 4} (1966) 83--105.

\bibitem{Wald74}
R.~M. Wald, {\it Black hole in a uniform magnetic field},  {\em Phys. Rev. {\rm
  D}} {\bf 10} (1974) 1680--1685.

\bibitem{Stephanibook}
H.~Stephani, D.~Kramer, M.~MacCallum, C.~Hoenselaers, and E.~Herlt, {\em Exact
  Solutions of Einstein's Field Equations}.
\newblock Cambridge University Press, Cambridge, second~ed., 2003.

\bibitem{AliFro04}
A.~N. Aliev and V.~P. Frolov, {\it Five-dimensional rotating black hole in a
  uniform magnetic field: {T}he gyromagnetic ratio},  {\em Phys. Rev. {\rm D}}
  {\bf 69} (2004) 084022.

\bibitem{EmpRea02prl}
R.~Emparan and H.~S. Reall, {\it A rotating black ring solution in five
  dimensions},  {\em Phys. Rev. Lett.} {\bf 88} (2002) 101101.

\bibitem{MisIgu06}
T.~Mishima and H.~Iguchi, {\it New axisymmetric stationary solutions of
  five-dimensional vacuum {E}instein equations with asymptotic flatness},  {\em
  Phys. Rev. {\rm D}} {\bf 73} (2006) 044030.

\bibitem{Figueras05}
P.~Figueras, {\it A black ring with a rotating 2-sphere},  {\em JHEP} {\bf 07}
  (2005) 039.

\bibitem{CohWal72}
J.~M. Cohen and R.~M. Wald, {\it Note on the angular momentum and mass of
  gravitational geons},  {\em J. Math. Phys.} {\bf 13} (1972) 543--545.

\bibitem{Herdeiro00}
C.~A.~R. Herdeiro, {\it Special properties of five-dimensional {BPS} rotating
  black holes},  {\em Nucl. Phys. {\rm B}} {\bf 582} (2000) 363--392.

\bibitem{Carter68}
B.~Carter, {\it Global structure of the {K}err family of gravitational fields},
   {\em Phys. Rev.} {\bf 174} (1968) 1559--1571.

\bibitem{ReiTre75}
C.~Reina and A.~Treves, {\it Gyromagnetic ratio of {E}instein-{M}axwell
  fields},  {\em Phys. Rev. {\rm D}} {\bf 11} (1975) 3031--3032.

\bibitem{CohTioWal73}
J.~M. Cohen, J.~Tiomno, and R.~M. Wald, {\it Gyromagnetic ratio of a massive
  body},  {\em Phys. Rev. {\rm D}} {\bf 7} (1973) 998--1001.

\bibitem{Aliev06prd}
A.~N. Aliev, {\it Rotating black holes in higher dimensional
  {E}instein-{M}axwell gravity},  {\em Phys. Rev. {\rm D}} {\bf 74} (2006)
  024011.

\bibitem{Aliev06}
A.~N. Aliev, {\it A slowly rotating charged black hole in five dimensions},
  {\em Mod. Phys. Lett.~{\rm A}} {\bf 21} (2006) 751--757.

\bibitem{Breckenridgeetal97}
J.~C. Breckenridge, R.~C. Myers, A.~W. Peet, and C.~Vafa, {\it {D}-branes and
  spinning black holes},  {\em Phys. Lett. {\rm B}} {\bf 391} (1997) 93--98.

\bibitem{Elvangetal05}
H.~Elvang, R.~Emparan, D.~Mateos, and H.~S. Reall, {\it Supersymmetric black
  rings and three-charge supertubes},  {\em Phys. Rev. {\rm D}} {\bf 71} (2005)
  024033.

\bibitem{KunNavVie06}
J.~Kunz, F.~Navarro-L\'erida, and J.~Viebahn, {\it Charged rotating black holes
  in odd dimensions},  {\em Phys. Lett. {\rm B}} {\bf 639} (2006) 362--367.

\bibitem{Emparan04}
R.~Emparan, {\it Rotating circular strings, and infinite non-uniqueness of
  black rings},  {\em JHEP} {\bf 03} (2004) 064.

\bibitem{PraPra05}
V.~Pravda and A.~Pravdov\'a, {\it {WAND}s of the black ring},  {\em Gen. Rel.
  Grav.} {\bf 37} (2005) 1277--1287.

\bibitem{Elvang03}
H.~Elvang, {\it A charged rotating black ring},  {\em Phys. Rev. {\rm D}} {\bf
  68} (2003) 124016.

\bibitem{Ortaggio05}
M.~Ortaggio, {\it Higher dimensional black holes in external magnetic fields},
  {\em JHEP} {\bf 05} (2005) 048.

\bibitem{IguMis06}
H.~Iguchi and T.~Mishima, {\it Solitonic generation of vacuum solutions in
  five-dimensional general relativity},  {\em Phys. Rev. {\rm D}} {\bf 74}
  (2006) 024029.

\bibitem{KinLasKun75}
A.~R. King, J.~P. Lasota, and W.~Kundt, {\it Black holes and magnetic fields},
  {\em Phys. Rev. {\rm D}} {\bf 12} (1975) 3037--3042.

\bibitem{BicJan85}
J.~Bi\v{c}\'ak and V.~Jani\v{s}, {\it Magnetic fluxes across black holes},
  {\em Mon. Not. R. Astron. Soc.} {\bf 212} (1985) 899--915.

\bibitem{BicDvo80}
J.~Bi\v{c}\'ak and L.~Dvo\v{r}\'ak, {\it Stationary electromagnetic fields
  around black holes. {III}. {G}eneral solutions and the fields of current
  loops near the {R}eissner-{N}ordstr{\"{o}}m black hole},  {\em Phys. Rev.
  {\rm D}} {\bf 22} (1980) 2933--2940.

\bibitem{ChaEmpGib98}
A.~Chamblin, R.~Emparan, and G.~W. Gibbons, {\it Superconducting $p$-branes and
  extremal black holes},  {\em Phys. Rev. {\rm D}} {\bf 58} (1998) 084009.

\bibitem{Mars:CQG1999}
M.~Mars, {\it A spacetime characterization of the {K}err metric},  {\em Class.
  Quantum Grav.} {\bf 16} (1999) 2507--2523.

\bibitem{Coleyetal04}
A.~Coley, R.~Milson, V.~Pravda, and A.~Pravdov\'a, {\it Classification of the
  {W}eyl tensor in higher dimensions},  {\em Class. Quantum Grav.} {\bf 21}
  (2004) L35--L41.

\bibitem{Milsonetal05}
R.~Milson, A.~Coley, V.~Pravda, and A.~Pravdov\'a, {\it Alignment and
  algebraically special tensors in {L}orentzian geometry},  {\em Int. J. Geom.
  Meth. Mod. Phys.} {\bf 2} (2005) 41--61.

\bibitem{Milson:2004wr}
R.~Milson, {\it Alignment and the classification of {L}orentz-signature
  tensors},  \href{http://xxx.lanl.gov/abs/gr-qc/0411036}{{\tt gr-qc/0411036}}.

\bibitem{Hall}
G.~S.~Hall, {\em Symmetries and Curvature Structure in General Relativity}.
\newblock World Scientific, Singapore, 2004.

\bibitem{ColPel06}
A.~Coley and N.~Pelavas, {\it Algebraic classification of higher dimensional
  spacetimes},  {\em Gen. Rel. Grav.} {\bf 38} (2006) 445--461.

\bibitem{FroSto03}
V.~P. Frolov and D.~Stojkovi\'c, {\it Particle and light motion in a space-time
  of a five-dimensional rotating black hole},  {\em Phys. Rev. {\rm D}} {\bf
  68} (2003) 064011.

\bibitem{Pravdaetal04}
V.~Pravda, A.~Pravdov\'a, A.~Coley, and R.~Milson, {\it Bianchi identities in
  higher dimensions},  {\em Class. Quantum Grav.} {\bf 21} (2004) 2873--2897.

\bibitem{BosEst81}
S.~K. Bose and E.~Esteban, {\it A null tetrad analysis of the {E}rnst metric},
  {\em J. Math. Phys.} {\bf 22} (1981) 3006--3008.

\bibitem{BenWar04}
I.~Bena and N.~P. Warner, {\it One ring to rule them all \ldots and in the
  darkness bind them?},  \href{http://xxx.lanl.gov/abs/hep-th/0408106}{{\tt
  hep-th/0408106}}.

\bibitem{GauGut05b}
J.~P. Gauntlett and J.~B. Gutowski, {\it General concentric black rings},  {\em
  Phys. Rev. {\rm D}} {\bf 71} (2005) 045002.

\end{thebibliography}

\providecommand{\href}[2]{#2}\begingroup\raggedright\endgroup

\end{document}